\documentclass[pra,aps,showpacs,tightenlines]{revtex4}
\usepackage{graphics,bm}
\usepackage{graphicx}
\usepackage{amsmath, amssymb, graphics}

\protect

\begin{document}
\title{Stability and dynamics of two-soliton molecules}
\author{U. Al Khawaja}
\affiliation{ \it Physics Department, United Arab Emirates University, P.O. Box
17551, Al-Ain, United Arab Emirates.}

\date{\today}

\begin{abstract}
The problem of soliton-soliton force is revisited. From the exact two solitons
solution of a nonautonomous Gross-Pitaevskii equation, we derive a generalized
formula for the mutual force between two solitons. The force is given for
arbitrary solitons amplitude difference, relative speed, phase, and separation.
The latter allows for the investigation of soliton molecules formation,
dynamics, and stability. We reveal the role of the time-dependent relative
phase between the solitons in binding them in a soliton molecule. We derive its
equilibrium bond length, spring constant, frequency, effective mass, and
binding energy of the molecule. We investigate the molecule's stability against
perturbations such as reflection from surfaces, scattering by barriers, and
collisions with other solitons.
\end{abstract}


\maketitle

\section{Introduction}
\label{intro_sec} The old-new interest in the problem of soliton-soliton
intertaction and soliton molecules has been increasingly accumulating
particularly over the past few years. This is mainly motivated by the
application of optical solitons as data carriers in optical fibers
\cite{solitons_books,soliton_book} and the realization of matter-wave solitons
in Bose-Einstein condensates \cite{randy,schreck}. One major problem limiting
the high-bit rate data transfer in optical fibers is the soliton-soliton
interaction. On the one hand, soliton-soliton interaction is considered as a
problem since it may destroy information coded by solitons sequences. On the
other hand, it is part of the problem's solution, since the interaction between
solitons leads to the formation of stable soliton molecules which can be used
as data carriers with larger ``alphabet'' \cite{mitschke1}.

The interaction force between solitons was first studied by Karpman and
Solov'ev using perturbation analysis \cite{solov}, Gordon who used the exact
two solitons solution \cite{gordon}, and Anderson and Lisak who employed a
variational approach \cite{anderson}. It was shown that the force of
interaction decays exponentially with the separation between the solitons and
depends on the phase difference between them such that in-phase solitons
attract and out-of-phase solitons repel. This feature was demonstrated
experimentally in matter-wave solitons of attractive Bose-Einstein condensates
\cite{randy,schreck} where a variational approach accounted for this repulsion
and showed that, in spite of the attractive interatomic interaction, the phase
difference between neighboring solitons indeed causes their repulsion
\cite{usama_randy}.

For shorter separations between the solitons, Malomed \cite{malomed} used a
perturbation approach to show that stationary solutions in the form of bound
states of two solitons are possible. However, detailed numerical analysis
showed that such bound states are unstable \cite{afan}. Stable bound states
were then discovered by Akhmediev \cite{akhmed} and a mechanism of creating
robust three-dimensional soliton molecules was suggested by Crasovan {\it et
al.} \cite{victor}. Recently, soliton molecules were realized experimentally by
Stratmann {\it et al.} in dispersion-managed optical fibers \cite{mitschke1}
and their phase structure was also measured \cite{mitschke2}. Perurbative
analysis was used to account theoretically for the binding mechanism and the
molecule's main features \cite{mitschke3,olivier}. Quantization of the binding
energy was also predicted numerically by Komarov {\it et al.} \cite{sanchez}.
In Refs.\cite{gardiner,lei}, a hamiltonian is constructed to describe the
interaction dynamics of solitons.

The mechanism by which the relative phase between the solitons leads to their
force of interaction, and hence the binding mechanism, is understood only
qualitatively as follows. For in-phase (out-of-phase) solitons, constructive
(destructive) interference takes place in the overlap region resulting in
enhancement (reduction) in the intensity. As a result, the attractive
intensity-dependent nonlinear interaction causes the solitons to attract
(repel) \cite{desem}. A more quantitative description is given in
Refs.~\cite{mitschke3,olivier}.

In view of its above-mentioned importance from the applications and fundamental
physics point of views, we address here the problems of soliton-soliton
interaction and soliton molecule formation using the exact two solitons
solution. This approach has been long pioneered by Gordon \cite{gordon} where
he used the exact two solitons solution of the homogeneous nonlinear
Schr$\rm\ddot o$dinger equation to derive a formula for the force of
interaction between two solitons, namely
\begin{equation}
{\ddot
\Delta}=-8\exp{(-\Delta)}\,\cos{\left(\Delta\phi\right)}\label{gordoneq},
\end{equation}
where $\Delta(t)$ is the solitons separation and $\Delta\phi(t)$ is their phase
difference. This formula was derived in the limit of large solitons separation
and for small difference in the center-of-mass speeds and intensities, which
limits its validity to slow collisions. With appropriately constructed
hamiltonian, Wu {\it et al.} have derived, essentially, a similar formula that
gives the force between two identical solitons and reliefs the condition on
slow collisions \cite{lei}.

Here, we present a more comprehensive treatment where we derive the force
between two solitons for arbitrary solitons intensities, center-of-mass speeds,
and separation. We also generalize Gordon's formula to inhomogeneous cases
corresponding to matter-wave bright solitons in attractive Bose-Einstein
condensates with time-dependent parabolic potentials \cite{randy,schreck} and
to optical solitons in graded-index waveguide amplifiers \cite{ragavan}. Many
interesting situations can thus be investigated. This includes the various
soliton-soliton collision regimes with arbitrary relative speeds, intensities,
and phases. Most importantly, soliton-soliton interaction at short solitons
separations will now be accounted for more quantitatively than before.
Specifically, soliton molecule formation is clearly shown to arise from the
time-dependence of the relative phase which plays the role of the restoring
force. In this case, the force between the two solitons is shown to be composed
of a part oscillating between attractive and repulsive, which arises from the
relative phase, and an attractive part that arises from the nonlinear
interaction. The time-dependence of the relative phase results in a natural
oscillation of the molecule's bond length around an equilibrium value. The
various features of the soliton molecule, including its equilibrium {\it bond
length}, {\it spring constant}, {\it frequency} and {\it amplitude} of
oscillation, and {\it effective mass}, will be derived in terms of the
fundamental parameters of the solitons, namely their intensities and the
nonlinear interaction strength.

The two solitons solution is derived here using the Inverse Scattering method
\cite{salle}. Although the two solitons solution of the homogeneous nonlinear
Schr$\rm\ddot o$dinger equation is readily known \cite{gordon,desem}, here we
not only generalize this solution to inhomogeneous cases, but also present it
in a new form that facilitates its analysis. The solution will be given in
terms of the four fundamental parameters of each soliton, namely the initial
amplitude, center-of-mass position and speed, and phase. The main features of
the solution will be shown clearly such as the contribution of the nonlinear
interaction to the actual separation and phase difference between solitons
where it turns that the separation between the two solitons increases with
logarithm of the difference between the amplitudes of the two solitons.
Furthermore, the general statement that {\it a state of two equal solitons with
zero relative speed and finite separation does not exist as a stationary state
for the homogeneous nonlinear Schr$\rm\ddot o$dinger equation}, will be
transparently and rigorously proved.

Stability of soliton molecules is an important issue since, in real systems,
perturbations caused by various sources such as losses, Raman scattering,
higher order dispersion, and scattering from local impurities, tend to destroy
the molecules. To investigate the stability of the soliton molecules described
by our formalism, we have considered three situations. First, we studied the
reflection of the molecule from a hard wall and a softer one. While for the
hard wall the molecule preserves its molecular structure after reflection, it
generally breaks up for the softer ones due to energy losses at the interface.
Secondly, the scattering of the molecule by a potential barrier was also
investigated. We show that the molecular structure is maintained only for some
specific heights of the barrier. This suggests a quantization in the binding
energy as predicted by Komarov {\it et al.} \cite{sanchez}. The oscillation
period of the reflected molecule is noticed to be smaller than for the incident
one. In addition, the outcome of scattering depends on the phase of the
molecule's oscillation at the interface of the barrier. For instance, a
dramatic change in the scattering outcome takes place if the {\it coalescence}
point of the molecule lies exactly at the interface. In such a case, the
otherwise totally transmitting molecule will now split into reflecting and
tunneling solitons. Thirdly, we have considered the collision between a single
soliton with a stationary soliton molecule. The effects of different initial
speeds, amplitudes, and phases of the scatterer soliton were studied. It turns
out that for slower collisions, it is easier for the scatterer soliton to break
up the soliton molecule, while for fast collisions the scatterer soliton expels
and then replaces one of the solitons in the molecule. The phase of the
scatterer soliton plays also a crucial rule in preserving or breaking the bond
of the molecule, which can be used as {\it key} tool to code or uncode data in
the molecule.

The rest of the paper is organized as follows. In section~\ref{exact_sec}, we
use the Inverse-Scattering method to derive the two solitons solution of the
inhomogeneous nonlinear Schr$\rm\ddot o$dinger equation and present the
solution in the above-mentioned appealing form. The main features of the
solution will be discussed in subsection~\ref{mainsec}. The center-of-mass
positions and relative phases will be derived in subsections~\ref{com_sec} and
\ref{phases_sec}, respectively. The force between solitons will be derived in
section~\ref{force_sec} where Gordon's formula will be extracted as a special
case in subsection~\ref{gordon_sec} and our more general formula will be
derived in subsection~\ref{our_sec}. In subsection~\ref{num_sec}, we compare
our formula with the numerical calculation. In section~\ref{molsec}, we show
the possibility of forming soliton molecules, derive their main features in
subsection~\ref{formation_sec}, and investigate their stability in
subsection~\ref{stability_sec}. We end in section~\ref{conc_sec} with a summary
of results and conclusions. The details of the derivation of the two solitons
solution and the center-of-mass positions are relegated to Appendices A and B,
respectively.

\section{The exact two solitons solution}
\label{exact_sec}

Matter-wave solitons of trapped Bose-Einstein condensates and optical solitons
in optical fibers can be both described by the dimensionless Gross-Pitaevskii
equation
\begin{equation}
i {\partial\over\partial t}\Psi(x,t)+{1\over2}{\partial^2\over\partial
x^2}\Psi(x,t)+ {1\over2}\left(
{\dot\gamma}(t)^2-{\ddot\gamma}(t)\right)x^2\Psi(x,t) +g_0\,e^{\gamma (t)}
|\Psi(x,t)|^2\Psi(x,t)=0 \label{gp},
\end{equation}
where $\gamma(t)$ is a dimensionless arbitrary real function. For matter-wave
solitons, length is scaled to the characteristic length of the harmonic
potential, $a_x=\sqrt{\hbar/m\omega_x}$, time to $1/\omega_x$, and the
wavefunction $\Psi(x,t)$ to $1/\sqrt{2a_x\omega_\perp/\omega_x}$, where
$\omega_x$ and $\omega_\perp$ are the characteristic frequencies of the quasi
one-dimensional ($\omega_\perp\gg\omega_x$) trapping potential  in the axial
and radial directions, respectively. In these units, the strength of the
interatomic interaction will be given by the ratio $g_0=a_s/a_x$, where $a_s$
is the $s$-wave scattering length. For the case of optical solitons, the
function $\Psi(x,t)$ represents the beam envelope, $t$ is the propagation
distance, $x$ is the radial direction, and the intensity-dependent term
represents the Kerr nonlinearity. In this case, scaling  is in terms of the
characteristic parameters of the fiber, as for instance in Ref.~\cite{lei}. The
specific form of the prefactors of the inhomogeneous and nonlinear terms
guarantees the integrability of this equation \cite{serk,usama_vib}. For the
special case of $\gamma(t)=0$, the homogeneous case is retrieved. Other
interesting special cases have also been considered
\cite{liang,serk,usama_vib}.

As outlined in Appendix~\ref{appa}, we use the Darboux transformation method to
derive the two solitons solution of this Gross-Pitaevskii equation, which can
be put in the form
\begin{eqnarray}\Psi(x,t)&=& \sqrt{\frac{n_1
\alpha_{11}(t)}{2}} e^{i (\phi_{01} + \phi_1(x,t))}\text{
}{\rm{sech}}[\alpha_{11}(t) (x-x_{\rm cm1}(t))]\nonumber\\
&+& \sqrt{\frac{n_2 \alpha_{22}(t)}{2}}e^{i (\phi_{01}+\phi_{02}+\phi_2(x,t)+
\tan^{-1}({\alpha_2\over\alpha_1})
+\tan^{-1}({\alpha_4\over\alpha_3}))}\nonumber\\
&\times&{\rm{sech}}\left[\alpha_{22}(t)(x-x_{\rm cm2}(t))
+\frac{1}{2}\log\left(\frac{\alpha_1^2+\alpha_2^2}{\alpha_3^2
+\alpha_4^2}\right)\right]\label{exactsol},
\end{eqnarray}
where\\\\
\begin{math}x_{\rm cmj}(t)=e^{-\gamma_0-\gamma (t)}\left( x_{j} e^{2 \gamma_0}+ g(t)\eta_{j}\right)\end{math},
$\hspace{1cm}j=1, 2$,\\\\
\begin{math}v_{\rm cmj}(t)={\dot x}_{\rm cmj}(t)\end{math},\\\\
\begin{math}\phi_j(x,t)=\frac{1}{8}g(t) \left(4 e^{-2 \gamma_0} \eta_j^2
+n_j^2 g_0^2\right)-\frac{1}{2} x_{\rm cmj}(t)^2 {\dot\gamma}(t)
+ v_{\rm cmj}(t) (x-x_{\rm cmj}(t))-\frac{1}{2} (x-x_{\rm cmj}(t))^2 {\dot\gamma}(t)\end{math},\\\\
\begin{math}\alpha_{jj}(t)={1\over2}n_jg_0 e^{\gamma (t)}\end{math},\\\\
\begin{math}\alpha_1=f_1+e^{y_m} \cos{z}\end{math},\\\\
\begin{math}\alpha_2=f_2+e^{y_m} \sin{z}\end{math},\\\\
\begin{math}\alpha_3=\frac{f_3}{2n_1 g_0}-2 n_1 g_0e^{y_p}\cos{z}\end{math},\\\\
\begin{math}\alpha_4=\frac{f_2}{2n_1 g_0}-2n_1 g_0 e^{y_p}\sin{z} \end{math},\\\\
\begin{math}{f_1}=(n_2+n_1) g_0+(n_2-n_1) g_0e^y\end{math},\\\\
\begin{math}{f_2}=2 (\eta_2-\eta_1)e^{-\gamma_0} \left(1+e^y\right)\end{math},\\\\
\begin{math}{f_3}=-(n_2-n_1) g_0-(n_2+n_1) g_0e^y\end{math},\\\\
\begin{math}y_m=\frac{1}{2}e^{\gamma (t)} g_0((n_1-n_2)x-(n_1 x_{\rm cm1}(t)-n_2 x_{\rm cm2}(t)))\end{math},\\\\
\begin{math}y_p=\frac{1}{2}e^{\gamma (t)} g_0((n_1+n_2)x-(n_1 x_{\rm cm1}(t)+n_2 x_{\rm cm2}(t)))\end{math},\\\\
\begin{math}y=e^{\gamma (t)} n_1 g_0 (x-x_{\rm cm1}(t))\end{math},\\\\
\begin{math}z_{jj}=-x_j \eta_j
+e^{-2 \gamma_0} \eta_j^2 g(0) + e^{-\gamma_0+\gamma (t)} x \eta_j +\frac{1}{8}
e^{-2 \gamma_0}g(t) \left(-4 \eta_j^2
+e^{2 \gamma_0} n_j^2 g_0^2\right)\end{math},\\\\
\begin{math}z=-\phi_{02}+z_{11}-z_{22}\end{math},\\\\
\begin{math}\eta_j=v_j+x_j {\dot\gamma}(0)\end{math},\\\\
\begin{math}g(t)=\int_0^te^{2 \gamma (t^\prime)}d\,t^\prime\end{math}.\\\\
The solution is put in this suggestive form to facilitate its analysis. The
first sech part corresponds to the exact single soliton solution with
center-of-mass position $x_{\rm cm1}(t)$, width $1/\alpha_{11}(t)$, phase
$\phi_{01}+\phi_1(x,t)$, and normalization $n_1$. Hence, $x_1$ and $v_1$
correspond to the initial center-of-mass position and speed, respectively. The
second sech term contains the same features in addition to a shift in both the
center-of-mass position and phase. It should be noted, however, that $x_{\rm
cmj}(t)$, $v_{\rm cmj}(t)$, $n_j$, and $\phi_j(x,t)$ correspond to the
center-of-mass position and speed, normalization, and phase of the single
noninteracting solitons. Due to the interaction between solitons, these four
characteristic quantities may not correspond exactly to the values of the same
physical quantities as they did for the single soliton solution. For instance,
$x_{\rm cm1}(t)$ will not correspond to the center-of-mass of one of the
solitons. Instead, the soliton may be shifted from that position due to the
interaction with the other soliton. In the following, we present a detailed
analysis of the locations and phases of the two solitons.

\subsection{Main features of the solution}
\label{mainsec} Inspection shows that there are two main regimes for the two
solitons solution, namely the regime of resolved solitons and the regime of
overlapping solitons. In the former case, the center-of-mass concept is well
defined and analysis of the relative dynamics becomes feasible. The solitons
are considered resolved as long as the two main peaks are not overlapping,
which means that partial overlap may occur in this regime. The analysis in this
section assumes the resolved solitons regime.

\begin{figure}
\begin{center}
\includegraphics[width=15cm]{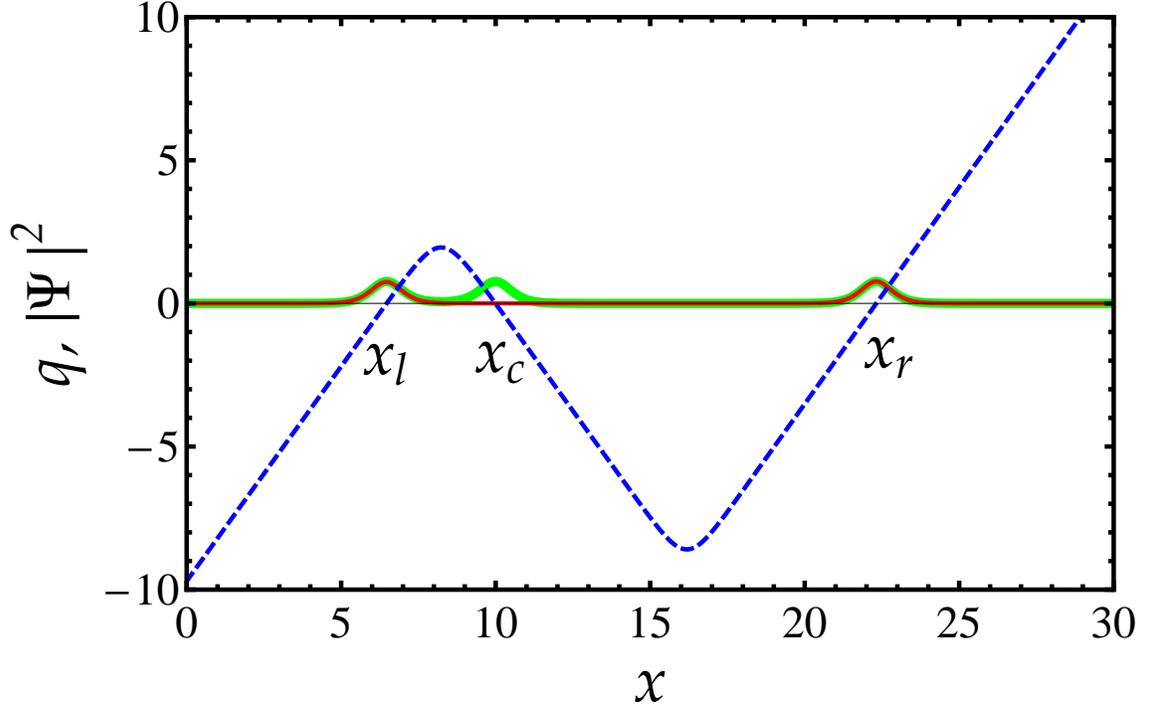}
\end{center}
\caption{(Color online) Dashed (blue) curve: The argument of the second sech
term in the two solitons solution, Eq.~(\ref{exactsol}),
$q=\alpha_{22}(t)(x-x_{\rm cm2}(t))
+(1/2)\log\left[(\alpha_1^2+\alpha_2^2)/(\alpha_3^2 +\alpha_4^2)\right]$. Thick
(green) curve: The soliton intensity $|\Psi^2(x,t)|^2$. Light (red) curve: The
soliton intensity with only the second sech term of Eq.~(\ref{exactsol}). The
parameters used are: $\gamma(t)=0$, $n_1=1$, $n_2=1.01$, $x_1=10$, $x_2=20$,
$v_1=v_2=0$, $g_0=3$, $\phi_{01}=\phi_{02}=0$, and $t=0$.} \label{fig1}
\end{figure}

In the resolved solitons regime, the argument of the second sech term of
Eq.~(\ref{exactsol}), namely $q=\alpha_{22}(t)(x-x_{\rm cm2}(t))
+(1/2)\log\left[(\alpha_1^2+\alpha_2^2)/(\alpha_3^2 +\alpha_4^2)\right]$,
simplifies to a function with three roots. The fact that the sech function is
peaked at the roots of its argument, leads to that the second sech term
corresponds to three ``solitons''. This is shown in Fig~\ref{fig1}. We denote
these solitons as the ``left'', ``central'', and ``right'' solitons with peak
locations at $x_l$, $x_c\approx x_{\rm cm1}(t)$, and $x_r$, respectively. We
notice that the central soliton is located at the position of the soliton of
the first sech term, namely near $x_{\rm cm1}(t)$. Further inspection shows
that the two solitons at this location are out of phase and interfere
destructively such that they do not appear in the total profile. Therefore, the
two solitons that our solution of Eq.~(\ref{exactsol}) describes are in fact
the left and right solitons arising from the second sech term. This is
different from what one aught to conclude from the form of the exact solution,
namely that the first sech term corresponds to one soliton and the second sech
term corresponds to the other soliton.

In general, the center-of-mass locations of the left and right solitons, $x_l$
and $x_r$, do not match $x_{\rm cm1}(t)$ and $x_{\rm cm2}(t)$. Typically, $x_l$
will be shifted to the left of $x_{\rm cm1}(t)$, $x_r$ will be shifted to the
right of $x_{\rm cm2}(t)$, while $x_c$ remains near $x_{\rm cm1}(t)$. The
amount of shift depends mainly on the normalization difference $n_2-n_1$ and
the relative speed $v_2-v_1$, as will be shown in the next subsection. An
interesting general and exact result is that, for the homogeneous case
$\gamma(t)=0$, a state of two equal solitons, $n_1=n_2$, with zero relative
speed, $v_1=v_2$, and finite separation, does not exist as an exact solution of
the Gross-pitaevskii equation. This can be proven by substituting  $n_1=n_2$
and $v_1=v_2$ in $q$ to find that the right and left solitons migrate to
$\infty$ and $-\infty$, respectively, while the center-of-mass of the central
soliton matches exactly $x_{\rm cm1}(t)$. Furthermore, we show in section
\ref{phases_sec} that the phase difference between this central soliton and
that of the first sech term equals, in this case, $\pi$; guaranteeing their
destructive interference. Thus, the three solitons disappear in such a special
case and $\Psi(x,t)$ becomes the trivial solution.

In view of the above, a need arises to derive formulae for the center-of-mass
positions, $x_l$ and $x_r$ in terms of the solitons parameters, which will then
be used to derive the force between the two solitons.

\subsection{Center-of-mass positions}
\label{com_sec} In this section, we derive formulae for the three roots of $q$
which correspond to the locations of the left, central, and right solitons,
$x_l$, $x_c$, and $x_r$, respectively. To facilitate the derivation, we define
$X=\exp{(y_m)}=\exp{\left[\frac{1}{2}e^{\gamma (t)} g_0\left((n_1-n_2)x-(n_1\,
x_{\rm cm1}(t)-n_2\, x_{\rm cm2}(t))\right)\right]}$,
$Y=\exp{(y)}=\exp{\left[e^{\gamma (t)} n_1 g_0 (x-x_{\rm cm1}(t))\right]}$,
$n_d=n_2-n_1$, $n_s=n_1+n_2$, $\eta_d=\eta_2-\eta_1$, and
$\eta_s=\eta_1+\eta_2$. The equation $q(X,Y)=0$ is a third-order polynomial in
both $X$ and $Y$. In principle, this equation can be solved algebraically for
$X$ or $Y$. However, extracting $x$ from the resulting three roots will not be
possible analytically for general $n_1\neq n_2$. Alternatively, and as can be
seen from Fig.~\ref{fig1}, we can exploit the simple linear behavior of $q$
near its roots.

Assuming, without loss of generality, $x_{\rm cm2}(t)>x_{\rm cm1}(t)$, and
noting that in the resolved solitons regime the solitons separation $|x_{\rm
cm2}(t)-x_{\rm cm1}(t)|$ is large, we argue in Appendix~\ref{appb1} that
$X\gg1$ for all $x$, $Y\gg1$ for $x\sim x_r$, $Y\sim1$ for $x\sim x_c$, and
$Y\ll1$ for $x\sim x_l$. Based on this, the center-of-mass position $x_r$ can
be derived from a Taylor expansion of $q$ in powers of large $X$ and $Y$.
Expanding $q$ in powers of large $X$ only and leaving $Y$ arbitrary, accounts
for $x_c$ and $x_l$ simultaneously.  Keeping terms up to first order of $1/X$,
we show in Appendix~\ref{appb1} that the position of the right soliton is given
by
\begin{equation}
x_r=x_{\rm cm2}(t)-\frac{2 e^{-\gamma (t)}
   \log{Y_r}}{ g_0 ({n_d}+{n_s})}
   +\frac{4 {Y_r}^{\frac{{n_s}}{{n_d}+{n_s}}}
 \left(g_0 {n_s}
   e^{\gamma_0} \cos{z_r}-2 {\eta_d} \sin{z_r}\right)e^{-\gamma (t)+\gamma_0}}{g_0 ({n_d}+{n_s})
   \left(g_0^2 {n_d}^2 e^{2 \gamma_0}
   +4 {\eta_d}^2\right)}\,e^{-\frac{1}{4} g_0 {x_d}
   ({n_s}-{n_d}) e^{\gamma (t)}}\label{xreq2},
\end{equation}
where
\begin{equation}
Y_r=\frac{g_0^2 ({n_d}-{n_s})^2 \left(g_0^2 {n_d}^2 e^{2 \gamma_0}+4
{\eta_d}^2\right)}{g_0^2 {n_s}^2
   e^{2 \gamma_0}+4 {\eta_d}^2}\label{beta5},
\end{equation}
and the position of the the left soliton is given by
\begin{equation}
x_l={x_{\rm cm1}}(t)-\frac{2 e^{-\gamma (t)} \log{Y_+}}
   {g_0 ({n_d}-{n_s})}-\frac{4 {C_+} \,e^{-\gamma (t)}}{
\sqrt{g_0^2 e^{2 \gamma_0} \left({n_d}^2-6 {n_d} {n_s}+{n_s}^2\right)-32
{\eta_d}^2}}e^{-\frac{1}{4} g_0 {x_d} ({n_s}+{n_d})
   e^{\gamma (t)}}
   \label{xleq2},
\end{equation}
where $x_d(t)=x_{\rm cm2}(t)-x_{\rm cm1}(t)$, $z_r=z(x_r)$, and $C_+$ and $Y_+$
are given in Appendix \ref{appb1}.

The second and third terms on the right hand side of Eqs.~(\ref{xreq2}) and
(\ref{xleq2}) account for the shift in the center-of-mass position with respect
to the single soliton ones. The third terms are much smaller than the second
ones since they decay exponentially with the solitons distance $x_d$. In the
limit $n_d\rightarrow0$ and $\eta_d\rightarrow0$, both $\log{Y_r}$ and
$\log{Y_+}$ take the form
$\log{[(n_d^2+4e^{-2\gamma_0}\eta_d^2/g_0^2)/n_s^2]}$. Thus, it is obvious that
for $n_d=\eta_d=0$, $x_r=\infty$ and $x_l=-\infty$. This agrees with our
earlier result that two equal solitons with zero relative speed and finite
separation, do not exist as an exact solution.

\subsection{Relative Phases}
\label{phases_sec}

In this section, we calculate the phases of the left, central, and right
solitons in reference to the phase of the soliton of the first sech part in the
two solitons solution. For simplicity, the special case of $n_d\ll1$ and
$\eta_d\ll1$ will be assumed. The phase difference between the left, central,
and right solitons on one hand and the soliton of the first sech term on the
other hand is generally given by
\begin{equation}
\Delta\phi=\phi_2(x,t)+\phi_{02}+{\rm tan}^{-1}({\alpha_2\over\alpha_1})+{\rm
tan}^{-1}({\alpha_4\over\alpha_3})-\phi_1(x,t),
\end{equation}
which by observing that for all $x$
\begin{equation}
\phi_2(x,t)+\phi_{02}-\phi_1(x,t)=z
\end{equation}
reduces to
\begin{equation}
\Delta\phi=z+{\rm tan}^{-1}({\alpha_2\over\alpha_1})+{\rm
tan}^{-1}({\alpha_4\over\alpha_3}).
\end{equation}
This expression gives the phases of the right soliton $\phi_r$, the central
soliton $\phi_c$, and the left soliton $\phi_l$, for $x\sim x_r$, $x_c$, and
$x_l$, respectively.

\begin{figure}
\begin{center}
\includegraphics[width=15cm]{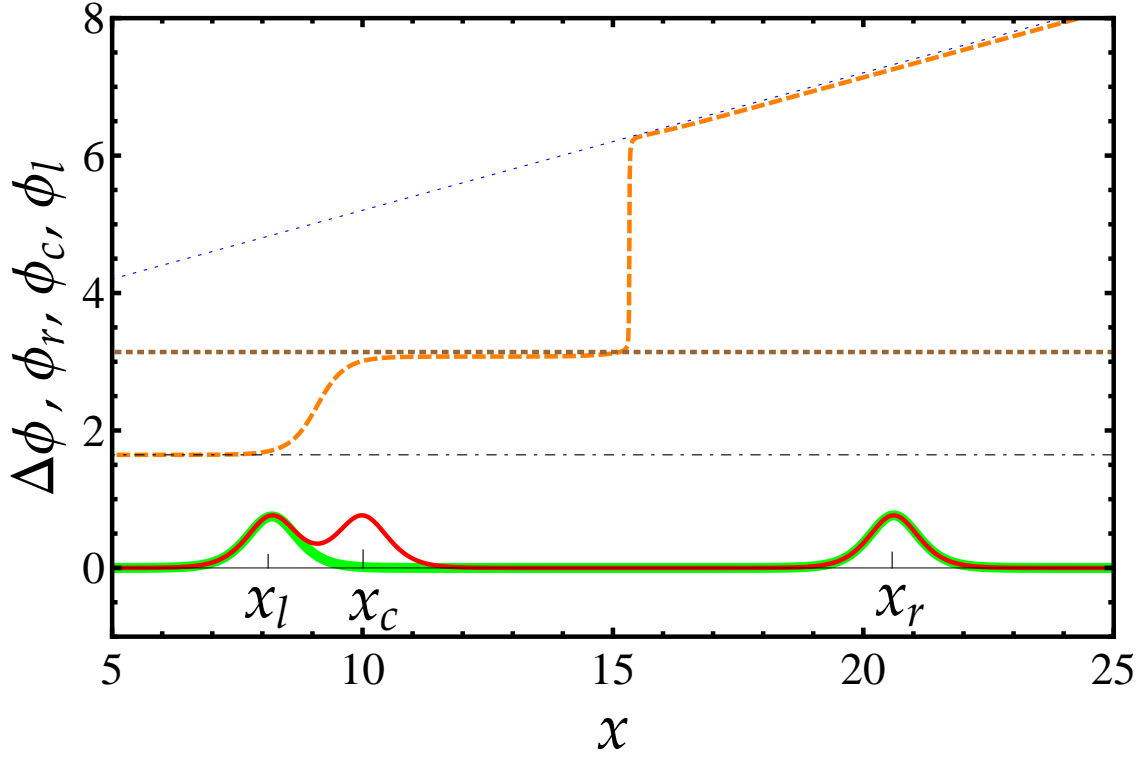}
\end{center}
\caption{(Color online) The phase of solitons (dashed curves) and solitons
intensity profiles (solid curves). Long dashed curve (orange) corresponds to
the exact phase of the solitons calculated directly from the two solitons
solution, Eq.~(\ref{exactsol}). Dotted (blue) curve corresponds to the phase of
the right soliton, $\phi_r$, calculated from Eq.~(\ref{phaser}). Thick dashed
curve corresponds to the phase of the central soliton, $\phi_c$, calculated
from Eq.~(\ref{phasec}). Dashed-dotted curve corresponds to the phase of the
left soliton, $\phi_l$, calculated from Eq.~(\ref{phasel}). The solid curves
correspond to the density profiles as in Fig.~\ref{fig1}. The parameters used
are: $\gamma(t)=0$, $n_1=1$, $n_2=1.01$, $x_1=10$, $x_2=20$, $v_1=1$,
$v_2=1.2$, $g_0=3$, $\phi_{01}=\phi_{02}=0$, and $t=0$.} \label{fig2}
\end{figure}

To calculate these phases we express the parameters $\alpha_{1-4}$ in terms of
$X$ and $Y$, as follows
\begin{equation}
\alpha_1=n_s\,g_0+n_d\,g_0\,Y+X\,\cos{z},
\end{equation}
\begin{equation}
\alpha_2=\eta_d\,e^{-\gamma_0}(1+Y)+X\,\sin{z},
\end{equation}
\begin{equation}
\alpha_3=-{n_d+n_s\,Y\over n_s-n_d}-(n_s-n_d)\,g_0\,{Y\over X}\,\cos{z},
\end{equation}
\begin{equation}
\alpha_4={2\eta_d\,e^{-\gamma_0}(1+Y)\over
(n_s-n_d)\,g_0}-(n_s-n_d)\,g_0\,{Y\over X}\,\sin{z}.
\end{equation}
In general, $X\gg1$ for all $x$, but $Y\gg1$ only for $x>x_{\rm cm1}(t)$, and
$Y=X\,e^{e^{\gamma(t)}g_0n_2(x-x_{\rm cm2}(t))/2}\gg X$ for $x>x_{\rm
cm2}(t)$, as shown in Appendix~\ref{appb1}.\\\\
{\bf I.\hspace{0.25cm} Phase of the right soliton $\phi_r$:}\\\\
In this case $x>x_{\rm cm2}(t)$ which leads to $Y\gg X$ and thus
$\alpha_1=n_d\,g_0\,Y$, $\alpha_2=2\eta_d\,e^{-\gamma_0}\,Y$,
$\alpha_3=-n_s\,Y/(n_s-n_d)$,
$\alpha_4=2\eta_d\,Y\,e^{-\gamma_0}/((n_s-n_d)g_0)$. Therefore\\\\\\
\begin{math}
\tan^{-1}{\alpha_2\over\alpha_1}=\tan^{-1}\left({2\eta_d\,e^{-\gamma_0}\over
g_0\,n_d}\right)=\left\{\begin{array}{ccc}
0,\hspace{1cm}&n_d\neq0,\eta_d=0,&\\\\
{\pi/2},\hspace{1cm}&n_d=0,\eta_d\neq0,&\\\\
\theta(n_d,\eta_d),\hspace{1cm}&n_d\neq0,\eta_d\neq0,&\\\\
\end{array}\right.
\end{math}\\\\\\
and
$\tan^{-1}{\alpha_4\over\alpha_3}=\tan^{-1}\left({-2\eta_d\,e^{-\gamma_0}\over
g_0\,n_s}\right)=\pi$, which gives
\begin{equation}
\phi_r=\pi+\theta-z\label{phaser},
\end{equation}
where $\theta$ is a function that depends on the ratio $\eta_d/n_d$ for nonzero
$\eta_d$ and $n_d$.
\\\\
{\bf II.\hspace{0.25cm} Phase of the central soliton $\phi_c$:}\\\\
In this case $x=x_{\rm cm1}(t)$ which gives $Y=1$, and hence
$\tan^{-1}{\alpha_2\over\alpha_1}=z$, and
$\tan^{-1}{\alpha_4\over\alpha_3}=\tan^{-1}\left({-1,0}\right)=\pi$. Finally,
we get
\begin{equation}
\phi_c=\pi\label{phasec}.
\end{equation}
The last result shows that, for $\eta_d=n_d=0$, i.e., two equal solitons with
zero relative speed, the central soliton and the soliton of the first sech term
of the two solitons solution are out of phase and therefore interfere
destructively.\\\\\\
{\bf III.\hspace{0.25cm} Phase of the left soliton $\phi_l$:}\\\\
In this case $x<x_{\rm cm1}(t)$ which results in $Y\ll1$ and
$\alpha_1=X\,\cos{z}$, $\alpha_2=X\,\sin{z}$, $\alpha_3=-n_d/(n_s-n_d)$,
$\alpha_4=2\eta_d\,e^{-\gamma_0}/(n_s-n_d)\,g_0$. Therefore,
$\tan^{-1}{\alpha_2\over\alpha_1}=z$ and
$\tan^{-1}{\alpha_4\over\alpha_3}=\tan^{-1}\left({-n_d\over 2\eta_d
e^{-\gamma_0}/g_0}\right)=\pi-\theta$, which gives
\begin{equation}
\phi_l=2\theta-z\label{phasel}.
\end{equation}
The phase difference between the left and right solitons is thus given by
\begin{math}
\Delta\phi=\phi_r-\phi_l=2\theta-z.
\end{math}
Noting that $\theta=0$ for $\eta_d=0$ and small but finite $n_d$, and
$\theta=\pi/2$ for $n_d=0$ and small but finite $\eta_d$, we finally conclude
that the phase difference between the two solitons is given by
\begin{equation}
\Delta\phi=\phi_r-\phi_l=\left\{\begin{array}{cc}\pi-z,\hspace{1cm}&n_d=0,\,
\eta_d\neq0,\\\\-z,\hspace{1cm}&\eta_d=0,\,
n_d\neq0.\end{array}\right.\label{phasediff}
\end{equation}
The above results are verified in Fig.~\ref{fig2}, where we plot $\Delta\phi$,
$\phi_r$, $\phi_l$, and $\phi_c$ versus $x$. The agreement between our
estimated values and the exact curve is evident.

\subsection{Soliton-soliton force}
\label{force_sec} In this section, we use the results of the previous two
subsections to derive the force between the left and right solitons. The force
is proportional to the acceleration of the solitons separation
\begin{eqnarray}
\Delta&=&x_r-x_l\nonumber\\&=&{x_d}(t)+\alpha
   e^{-\gamma (t)}\nonumber\\&+& \left[{\beta_1} \cos ({z_r}(t)) e^{\frac{1}{4}
   g_0 {n_d} {x_d}(t) e^{\gamma (t)}-\gamma (t)+2 \gamma_0}\right.\nonumber\\&+&{\beta_2}
\cos ({z_l}(t)) e^{-\frac{1}{4}
   g_0 {n_d} {x_d}(t) e^{\gamma (t)}-\gamma (t)}
   \nonumber\\&+&{\beta_3}
   \sin ({z_l}(t)) e^{-\frac{1}{4} g_0
   {n_d} {x_d}(t) e^{\gamma (t)}-\gamma (t)-\gamma_0}\nonumber\\
   &+&\left.{\beta_4} \sin ({z_r}(t))
   e^{\frac{1}{4} g_0 {n_d} {x_d}(t) e^{\gamma (t)}-\gamma (t)
   +\gamma_0}\right]\,e^{-\frac{1}{4} g_0 {n_s} {x_d}(t) e^{\gamma (t)}}\label{deltaeq},
\end{eqnarray} where
\begin{equation}
\alpha =\frac{2 (({n_s}-{n_d}) \log ({Y_r})+({n_d}+{n_s}) \log ({Y_+}))}{g_0
   \left({n_d}^2-{n_s}^2\right)},
\end{equation}
and the coefficients $\beta_{1-4}$ are given in Appendix~\ref{appb2}. The first
two terms on the right hand side of Eq.~(\ref{deltaeq}) are the dominant ones
since they correspond to the noninteracting solitons separation, $x_d(t)$, and
their logarithmic shifts $\propto\alpha$ arising from the interaction between
solitons. The time-dependence of $\Delta$ originates from $\gamma(t)$,
$z_{r,l}(t)$, and $x_d(t)$. The acceleration can thus be derived
\begin{equation}
{\ddot\Delta}(t)=
\left({\dot\gamma}(t)^2-{\ddot\gamma}(t)\right)\,\Delta(t)+\frac{1}{64}\,\left[{\beta_5}
\sin ({z_l}(t))+{\beta_6} \sin
   ({z_r}(t))+{\beta_7} \cos ({z_l}(t))+{\beta_8} \cos ({z_r}(t))\right]
   \, e^{-\frac{1}{4} g_0 n_s {\Delta}(t) e^{\gamma
(t)}+3 \gamma (t)}\label{force2},
\end{equation}
where the coefficients $\beta_{5-8}$ are given in Appendix~\ref{appb2}. In the
last equation, we have used Eq.~(\ref{deltaeq}) with only the first two terms
of its right hand side to substitute for $x_d(t)$ in terms of $\Delta(t)$ in
the exponential factor. The first term on the right hand side of
Eq.~(\ref{force2}) corresponds to the force due to the external potential which
vanishes for the homogeneous case. The rest of the terms correspond to the
force of interaction between the solitons. The interaction force depends, as
expected, on the phase difference of the two solitons and decays exponentially
with their separation. It should be noted that this equation is a
generalization of Gordon's formula \cite{gordon} in two aspects. First, it is
derived for a time-dependent inhomogeneous medium. Secondly we have,
essentially, no restriction on the difference  between the two solitons
amplitudes and speeds; apart from some extreme cases which were mentioned in
Appendix \ref{appb1} and will be discussed further below.

\subsubsection{Gordon's formula}
\label{gordon_sec} For the homogeneous case, $\gamma(t)=0$, and in the limits
$n_d\rightarrow0$ and $\eta_d\rightarrow0$, the acceleration formula,
Eq.~(\ref{force2}), simplifies considerably. An apparent inconsistency occurs
when switching the order of these two limits, namely
\begin{equation}
\lim_{n_d\rightarrow0}\,\lim_{\eta_d\rightarrow0}\,{\ddot\Delta}(t)
=-{1\over8}\,(n_s\,g_0)^3\,e^{-{1\over4}n_s\,g_0\,\Delta}\,\cos{z},
\end{equation}
while
\begin{equation}
\lim_{\eta_d\rightarrow0}\,\lim_{n_d\rightarrow0}\,{\ddot\Delta}(t)=
{1\over8}\,(n_s\,g_0)^3\,e^{-{1\over4}n_s\,g_0\,\Delta}\,\cos{z},
\end{equation}
which differ by an overall minus sign. The conflict is resolved by invoking
Eq.~(\ref{phasediff}) where it is shown that in the first case: $z=\Delta\phi$,
while in the second case: $z=\pi-\Delta\phi$. Therefore, the two approaches
agree on the following result
\begin{equation}
{\ddot\Delta}(t)
=-{1\over8}\,(n_s\,g_0)^3\,e^{-{1\over4}n_s\,g_0\,\Delta}\,\cos{\Delta\phi}\label{gordon2}.
\end{equation}
This is essentially Gordon's result, Eq.~(\ref{gordoneq}), since in his
derivation Gordon took $g_0=1$ and soliton amplitude $n_j\sqrt{g_0}/2=1$,
$j=1,2$, as can be seen in Eqs.(1) and (6) of Ref.~\cite{gordon}. Substituting
$g_0=1$ and $n_s=4$ in the last equation, it becomes identical to
Eq.~(\ref{gordoneq}). It should be mentioned here that Eq.~(\ref{gordoneq}) was
also derived in Ref.~\cite{solov} using a perturbation analysis based on the
Inverse Scattering method, and in Ref.~\cite{anderson} using a variational
calculation.

\subsubsection{Our formula}
\label{our_sec} For nonzero $\gamma(t)$ and in the limits $n_d\rightarrow0$ and
$\eta_d\rightarrow0$, the acceleration formula takes the form
\begin{equation}
{\ddot\Delta}(t) =\left({\dot\gamma}(t)^2-{\ddot\gamma}(t)\right)\,\Delta(t)
-\frac{1}{16}\,(g_0\, n_s)^3\, e^{-\frac{1}{4} g_0 n_s \Delta(t)\, e^{\gamma
(t)}+3\gamma(t)-\gamma_0}\,(1+e^{\gamma_0})\,\cos ({\Delta\phi}).
\end{equation}
This is a generalization to Gordon's formula for the inhomogeneous case as
modeled by Eq.~(\ref{gp}). Depending on the specific form of $\gamma(t)$, the
two force terms, namely the external (first term) and the interaction (second
term), can be repulsive, attractive, or oscillatory. In addition, the phase
difference $\Delta\phi(t)$ also depends on $\gamma(t)$. It is established in
the homogeneous case, as will also be shown in section \ref{molsec}, that the
time-dependence of the phase difference is responsible for binding the two
solitons in the soliton molecule. Here, in addition to the possibility of
forming soliton molecules, the dependence of $\Delta\phi(t)$ on $\gamma(t)$
allows for controlling the parameters of the molecule such as its equilibrium
bond length, period, and spring constant. This and possibly other interesting
phenomena will be left for future investigation.

\subsubsection{Comparison with numerical calculations}
\label{num_sec} 
\begin{figure}
\begin{center}
\includegraphics[width=15cm]{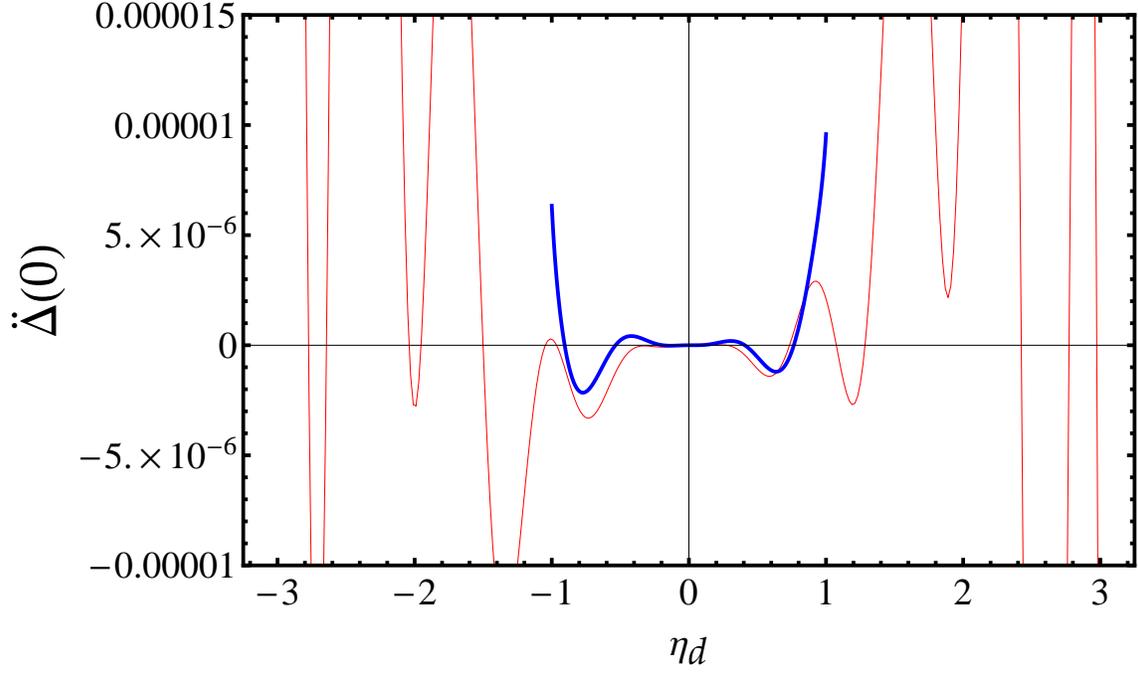}
\end{center}
\caption{(Color online) The initial acceleration of the solitons separation,
${\ddot\Delta}(0)$, versus $\eta_d$. Thick (blue) curve is calculated from
Eq.~(\ref{force2}). Light (red) curve is calculated numerically from the exact
two solitons solution, Eq.~(\ref{exactsol}). The parameters used are:
$\gamma(t)=0$, $n_1=1$, $n_2=1.01$, $x_1=10$, $x_2=20$, $v_1=1$, $g_0=3$,
$\phi_{01}=\phi_{02}=0$, and $t=0$.} \label{fig3}
\end{figure}

To obtain an estimate of the accuracy of the general
acceleration formula, Eq.~(\ref{force2}), we calculate numerically the
acceleration from the exact two solitons solution, Eq.~(\ref{exactsol}), and
compare the two results. The distance between the two solitons of the function
$|\Psi(x,t)|^2$ is determined using a numerical algorithm that employs our
formulae for $x_l$ and $x_r$ given by Eqs.~(\ref{xreq2}) and (\ref{xleq2}) to
calculate seed values. The distance is then differentiated numerically twice at
$t=0$. In Fig.~\ref{fig3}, we compare the two results. Good agreement is
obtained for $|\eta_d|\lesssim1$. The analytical solution diverges at
$\eta_d\thickapprox\pm1$. The value at which divergence takes place is set by
the specific choice of parameters in Fig.~\ref{fig3}. As pointed out in section
\ref{com_sec}, the divergence occurs due to the merging of the central and left
solitons. This artifact divergency can be remedied by associating the location
of the local maximum of $q$ to $x_l$ once this maximum has reached the $x$-axis
from above.

\begin{figure}
\begin{center}
\includegraphics[width=15cm]{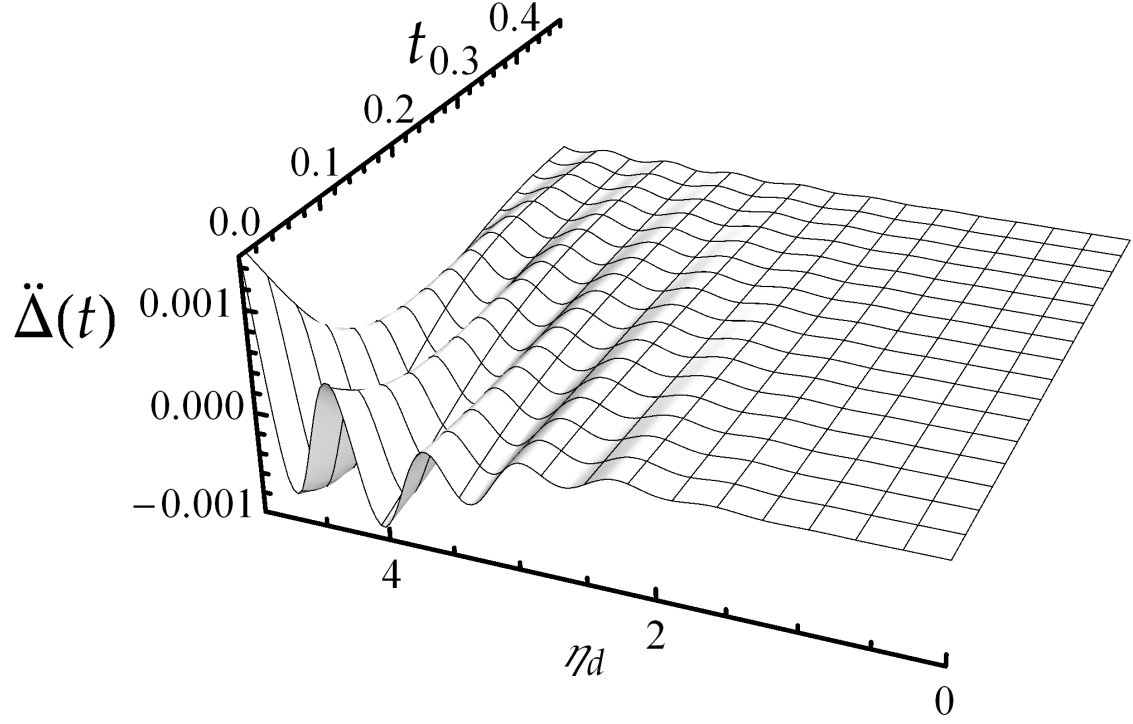}
\end{center}
\caption{The acceleration of the solitons separation, ${\ddot\Delta}(0)$,
versus $\eta_d$ and $t$. The parameters used are: $\gamma(t)=0$, $n_1=1$,
$n_2=1.01$, $x_1=10$, $x_2=20$, $v_1=1$, $g_0=3$, $\phi_{01}=\phi_{02}=0$.}
\label{fig4}
\end{figure}

Restricting our study to the region where agreement is obtained, we
interestingly notice that the acceleration is oscillating between positive and
negative values. This means that the force between the solitons is oscillating
between attractive and repulsive. The possibility of having attractive forces
for finite $\eta_d$ is particularly interesting; for two solitons with nonzero
relative positive speed, i.e. the solitons are initially diverging from each
other, the force between them is attractive. This suggests that, if the force
remains attractive for sufficient time, the two solitons will slow down and
eventually converge at some point. If true, this should occur at small
distances since the force decays exponentially with distance and when the two
solitons are allowed to diverge even for a short while, the force might be
weakened such that the two solitons can not return back. To be able to judge on
such a possibility, we need to know what happens to the acceleration at later
times. To that end, we calculate numerically the acceleration in terms of
$\eta_d$ and $t$. The result is plotted in Fig.~\ref{fig4}, where it is clear
that the acceleration indeed decays with time for all $\eta_d$. This leads to
that any nontrivial effect of the oscillating force is most likely to take
place at short solitons separations. This is what we find in the next section
where the possibility of forming stable soliton molecules is pointed out.

\section{Soliton molecule}
\label{molsec} We have shown in the previous section that, as a result of the
solitons time-dependent relative phase, the force of interaction between
solitons is oscillating between repulsive and attractive. Since the force
decays exponentially with the solitons separations, this oscillation will have
a tangible effect only when the two solitons are close to each other. In this
section, we investigate the force of interaction between solitons for short
solitons separation. In such a special case, Eq.~(\ref{deltaeq}) takes a simple
form that accounts for the solitons separation in terms of their relative
phase. Using this formula, we show that the solitons will be bound to oscillate
around some equilibrium distance where the phase plays the role of the
restoring force. Comparison with exact numerical calculations shows that this
formula is accurate for almost the full range of the solitons separation,
except at the coalescence point (if any). In subsection~\ref{formation_sec}, we
discuss the main features of the resulting soliton molecules, and in
subsection~\ref{stability_sec}, we investigate numerically their stability in
different scattering regimes.

To focus on the role of relative phase, we simplify the analysis by restricting
our treatment to the homogeneous case, $\gamma(t)=0$, and zero relative speed,
$\eta_d=v_d=v_2-v_1=0$. We also set $x_1=x_2$ so that any separation between
the solitons to be as small as possible which in this case arises only from the
logarithmic shifts ($\sim\alpha$ in Eq.~(\ref{deltaeq})). In this case, the
solitons separation $\Delta$, given by Eq.~(\ref{deltaeq}), simplifies in the
limit $n_d\ll n_s$ to
\begin{equation}
\Delta(t)=\frac{4}{g_0 n_s} \log \left[1+2\, n_s\,g_0\,\cos
{\left(\frac{1}{8}\,
   g_0^2\, n_d\, n_s\, t\right)+(n_s\,g_0)^2}\right]-\frac{4}{g_0 n_s}\log
\left(\frac{g_0
   n_d^2}{n_s}\right)\label{delmol},
\end{equation}
and the acceleration is given by
\begin{equation}
{\ddot\Delta}(t)=-{1\over8}\,(n_s\,g_0)^3\,\frac{2\,n_s\,g_0+ (n_s\,g_0)^2\,
\cos \left(\frac{1}{8}\,
   g_0^2\, n_d\, n_s\, t\right)
   +\cos \left(\frac{1}{8}\, g_0^2\, n_d\,
   n_s\, t\right)}{
   1+2\,n_s\, g_0\, \cos \left(\frac{1}{8}\,
   g_0^2\, n_d\, n_s\, t\right)+(n_s\,g_0)^2}\,e^{-\frac{1}{4} g_0 n_s {\Delta
\,x}}\label{delppmol}.
\end{equation}
It is noted that for $n_s\,g_0\gg1$ or $n_s\,g_0\ll1$, Gordon's formula is
retrieved, but here with an explicit time-dependence of the phase,
$\Delta\phi=n_d\,n_s\,g_0\,t/8$. This acceleration formula deviates
considerably from Gordon's formula for $n_s\,g_0\sim1$. Specifically, for
$n_s\,g_0=1$, $\Delta$ diverges to $-\infty$ at
$\cos{(n_d\,n_s\,g_0\,t/8)}=-1$, which indicates that the two solitons
coalesce. This is confirmed below by examining the exact solution at this
condition. We note here that an approximate expression for the solitons
separation was also derived in Refs.~\cite{solov,anderson,desem}. In addition,
our predicted molecule's oscillation frequency (see Eq.~(\ref{freqeq}) below)
agrees with these references.

\begin{figure}
\begin{center}
\includegraphics[width=15cm]{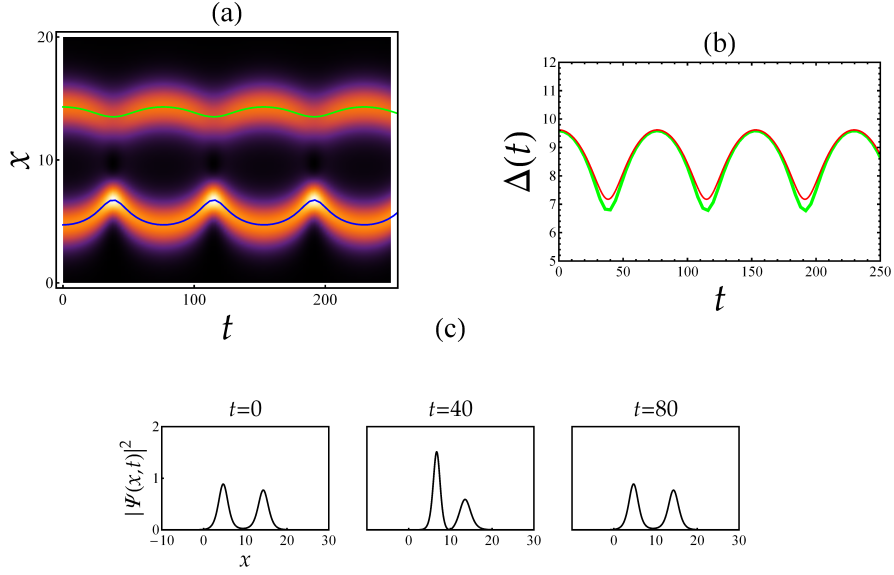}
\end{center}
\caption{(Color online) Density profile of the soliton molecule and the
solitons center-of-mass trajectories and separations. {\bf (a)} Density plot
corresponds to solitons density. Curves correspond to solitons trajectories
calculated numerically from the exact solution (\ref{exactsol}). {\bf (b)}
Thick (green) curve is solitons separation calculated from the exact solution
(\ref{exactsol}). Light (red) curve is the solitons separation calculated from
formula (\ref{delmol}). {\bf (c)} Density profile at some specific times. The
parameters used are: $\gamma(t)=0$, $n_1=2.37$, $n_d=0.5$, $x_1=x_2=10$,
$v_1=v_2=0$, $g_0=0.5$, $\phi_{01}=\phi_{02}=0$.} \label{fig5}
\end{figure}

\begin{figure}
\begin{center}
\includegraphics[width=15cm]{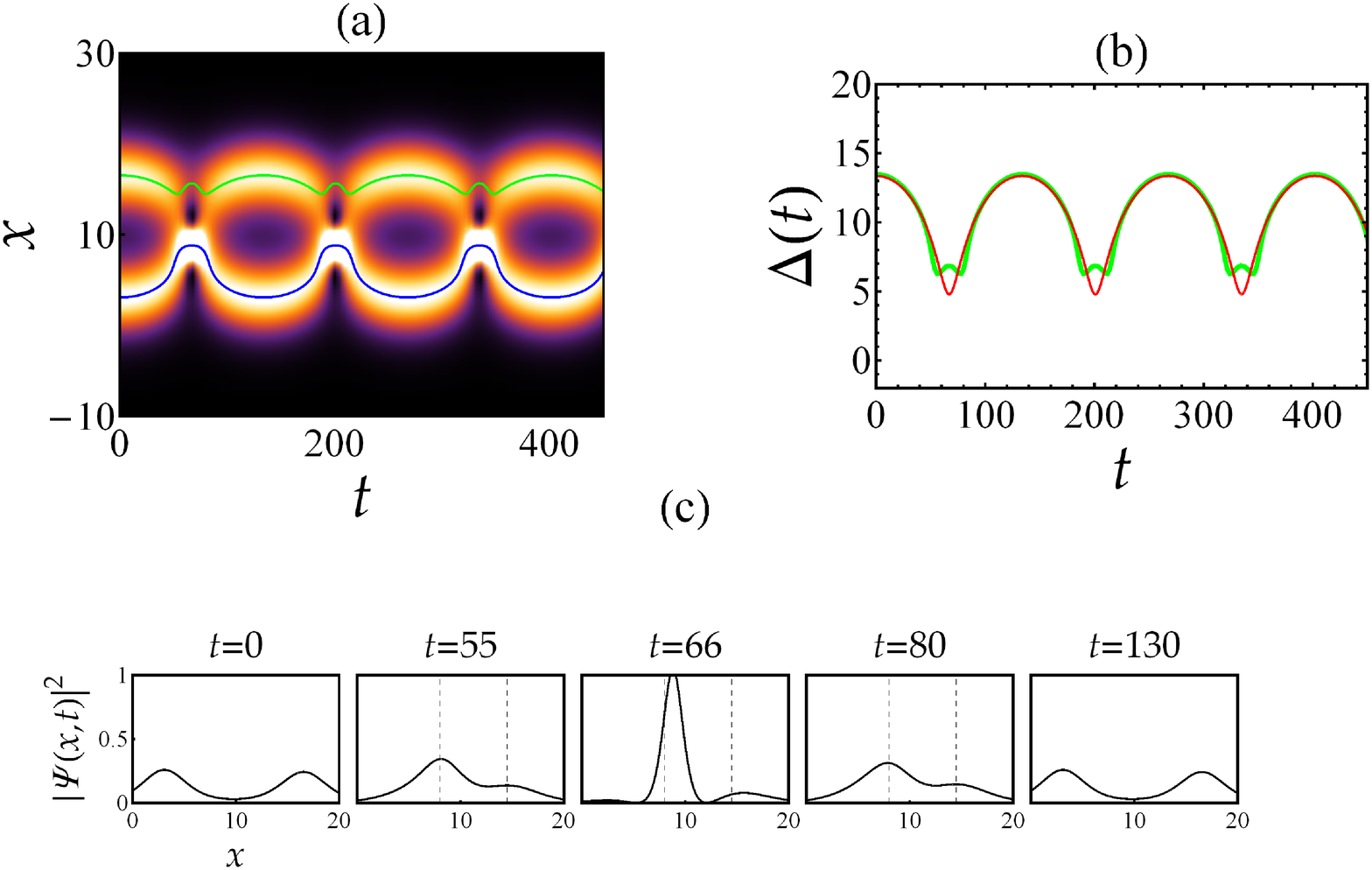}
\end{center}
\caption{Same as Fig.~\ref{fig5} but with $n_1=1.25$.} \label{fig6}
\end{figure}

\begin{figure}
\begin{center}
\includegraphics[width=15cm]{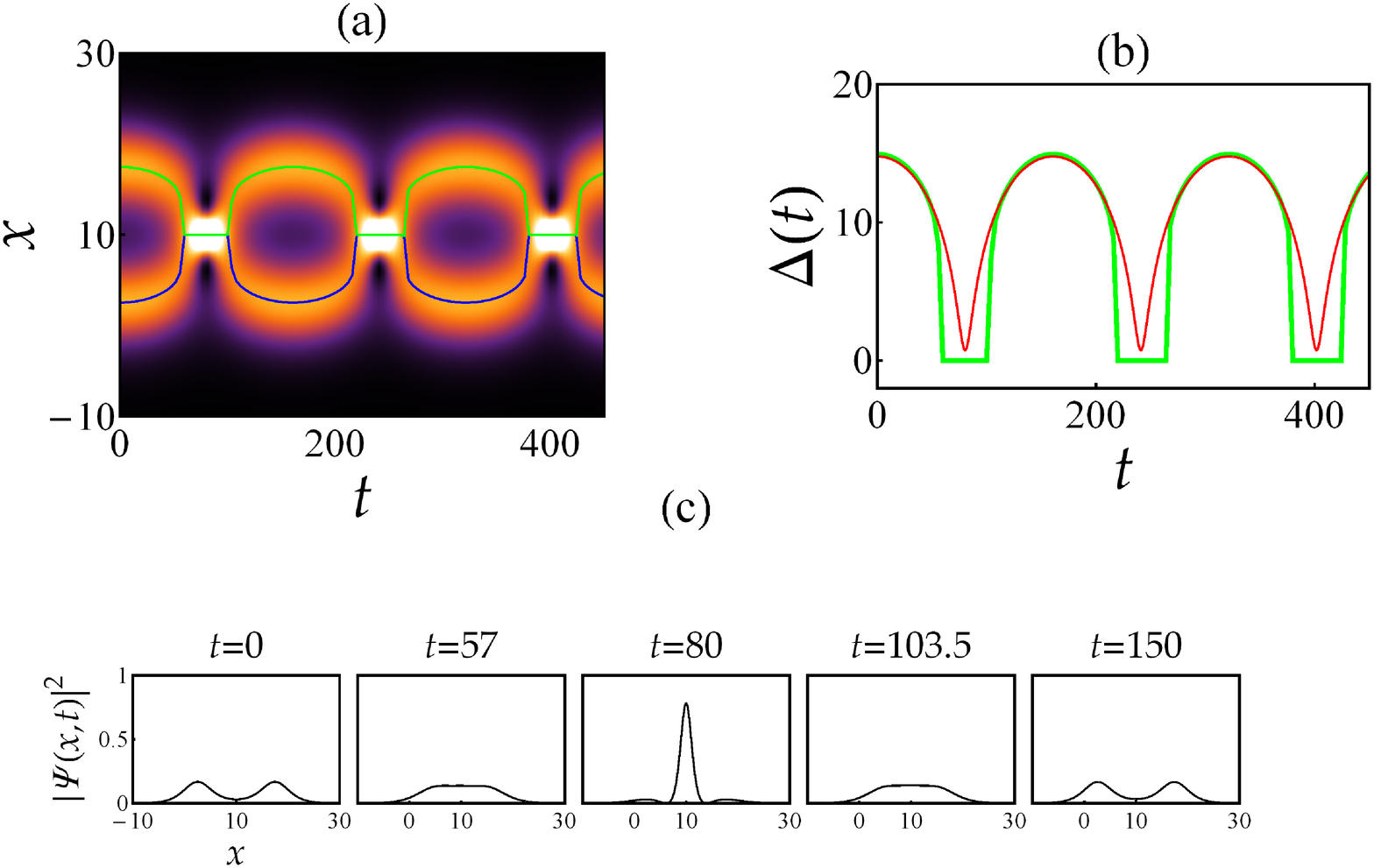}
\end{center}
\caption{Same as Fig.~\ref{fig5} but with $n_1=1$.} \label{fig7}
\end{figure}

To verify this feature, we calculate numerically the distance between the two
solitons directly from the exact solution, Eq.~(\ref{exactsol}), for different
values of $n_s\,g_0$. For $n_s\,g_0=2.5$, the density plot in Fig.~\ref{fig5}a,
shows a soliton molecule of two clearly resolved solitons with a separation
oscillating around some nonzero equilibrium distance. Approaching the solitons
coalescence point with $n_s\,g_0=1.5$, the density plot in Fig.~\ref{fig6}a,
shows the two solitons approaching each other more than the previous case.
Furthermore, this figure shows a slight bounce back by one of the solitons in
the region of collision. Approaching further the coalescence condition with
$n_s\,g_0=1.25$, we indeed observe in Fig.~\ref{fig7}a that the two solitons
merge almost completely. For a more quantitative comparison, we calculate
numerically the center-of-mass trajectories of the two solitons. We show the
trajectory curves in the density plots of Figs.~\ref{fig5}a-\ref{fig7}a. In
Figs.~\ref{fig5}b-\ref{fig7}b, we plot the solitons separation obtained from
formula (\ref{delmol}) and the numerical trajectories obtained from the exact
solution. It is clear from these figures that this formula agrees well with the
exact soliton separation except near the collision region. In Fig.~\ref{fig5}b,
the two solitons remain away from each other during the collision, and
therefore good agreement is obtained with the exact result even in the
collision region. In Fig.~\ref{fig6}b the two solitons approach each other
further such that formula (\ref{delmol}) does not account for the
above-mentioned slight bounce of one of the solitons. In Fig.~\ref{fig7}b,
agreement with the exact solution in the collision region is qualitative. We
found that at the condition $n_s\,g_0=1$ and for $n_d\ll n_s$, the analytic
curve overlaps with that of the exact solution; apart from the horizontal
segments where formula (\ref{delmol}) diverges to $-\infty$. Further insight is
obtained by plotting the density profile of the soliton molecule at some
specific times, as shown in Figs.~\ref{fig5}c-\ref{fig7}c. In Fig.~\ref{fig5}c,
we observe that the initial amplitude imbalance is never removed during the
dynamics. Instead, it becomes maximum when the two solitons are closest to each
other. In addition, we notice that the oscillation amplitude of the larger
soliton around its equilibrium position is larger. Figure~\ref{fig6}c shows
clearly the soliton bounce which takes place in the time interval $t=55$ to
$t=80$. In these subfigures we plot two vertical dashed lines that indicate the
position of the solitons at the closest approach. It is clear that after first
closest approach at $t=55$, the right soliton bounces back with a maximum
displacement at $t=66$. In Fig.~\ref{fig7}c, it is shown that, although the two
solitons coalesce, two small symmetric $\it wings$ appear. A detailed
examination of these wings shows that they are the remnants of the two solitons
after they coalesce and they both bounce back in the collision region similar
to the case of Fig.~\ref{fig6}.

\begin{figure}
\begin{center}
\includegraphics[width=15cm]{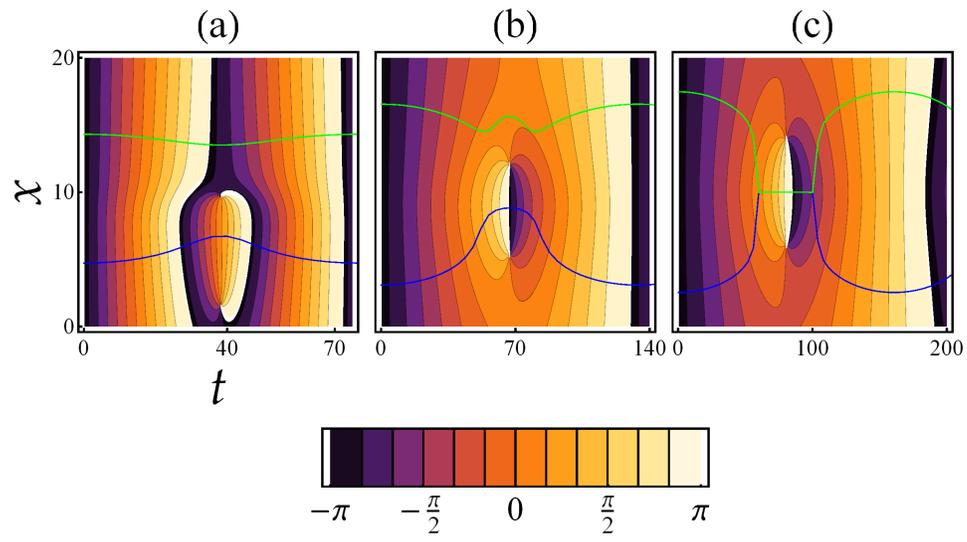}
\end{center}
\caption{Contour plots showing the phase of the soliton molecules of
Figs.~\ref{fig5}-\ref{fig7}. Subfigures {\bf (a)}, {\bf (b)}, and {\bf (c)},
correspond to Fig.~\ref{fig5}, Fig.~\ref{fig6}, and Fig.~\ref{fig7},
respectively. The blue (lower) and green (upper) curves correspond to the
center-of-mass trajectories of the left and right solitons, respectively.}
\label{fig8}
\end{figure}

It is also instructive to show the dynamics of the phase profile during the
molecule's oscillation. This is shown with the contour plots of Fig.~\ref{fig8}
which correspond to the molecules of Figs.~\ref{fig5}-\ref{fig7}. In
Fig.~\ref{fig8}a, which corresponds to Fig.~\ref{fig5}, the two solitons start
initially in phase. By time the phase of the right soliton, which is the one
with higher intensity amplitude and larger oscillation displacement, starts to
exceed that of the left soliton. At the point of closest approach, the phase
difference is exactly $\pi$. After that point, the two solitons diverge again,
the phase difference starts to decrease, and the cycle is repeated. Similar
behavior is seen in Fig.~\ref{fig8}b. However, in Fig.~\ref{fig8}c, where the
two solitons coalesce for a considerable amount of time, the phase difference
during the coalescence time is zero. It is thus not completely understood why,
in this case, the two solitons still repel each other and eventually split.

\subsection{Molecule formation and dynamics}
\label{formation_sec} Having established the existence of the soliton molecule
from the exact two solitons solution and derived a formula that describes its
bond length, here we use this formula to examine more closely the propertie of
the soliton molecule and its mechanism of binding.

It is clear from Eq.~(\ref{delppmol}) that the sinusoidal time-dependence of
the solitons relative phase leads to a force of interaction that oscillates
between attractive and repulsive and hence allowing for soliton molecule
formation. Further details of the mechanism of binding will be uncovered by
expressing the acceleration, ${\ddot\Delta}(t)$ in terms of $\Delta(t)$ by
substituting for $\cos{(n_s\,n_d\,g_0^2\,t/8)}$ from Eq.~(\ref{delmol}) into
Eq.~(\ref{delppmol}), to get
\begin{equation}
{\ddot\Delta}=\frac{1}{16} g_0 n_s \left(1-g_0^2 n_s^2\right){}^2
   \left(\frac{n_s}{n_d}\right)^2 e^{-\frac{1}{2} g_0 n_s {\Delta}}-\frac{1}{16}
   \left(g_0 n_s\right){}^2 \left(g_0^2 n_s^2+1\right) e^{-\frac{1}{4} g_0 n_s
   {\Delta}}\label{delpp2mol}.
\end{equation}
This shows that the interaction force between the two solitons is the resultant
of an attractive part and a repulsive part. The equilibrium bond length,
defined by ${\ddot\Delta}(\Delta=\Delta_{\rm eq})=0$, is given by
\begin{equation}
{\Delta_{\rm eq}}=\frac{4}{g_0 n_s} \log \left(\frac{n_s
   \left(g_0^2 n_s^2-1\right){}^2}{{n_d^2}\,g_0 \left(g_0^2
n_s^2+1\right)}\right).
\end{equation}
In consistence with our previous result, the equilibrium bond length diverges
as $-\log{n_d}$. Solving the last equation for $n_d^2$ and then substituting in
Eq.~(\ref{delpp2mol}), ${\ddot\Delta}$ simplifies to
\begin{equation}
{\ddot\Delta}=\frac{1}{16} g_0^2 n_s^2 \left(g_0^2 n_s^2+1\right)
\left[\left(e^{-\frac{1}{4}
   g_0 n_s \left({\Delta}-\frac{{\Delta_{\rm eq}}}
   {2}\right)}\right){}^2-e^{-\frac{1}{4} g_0
   n_s {\Delta}}\right].
\end{equation}
For small amplitude oscillations, $\Delta\simeq\Delta_{\rm eq}$,  the last
equation gives
\begin{equation}
{\ddot\Delta}=-\frac{1}{64} g_0^3 n_s^3 \left(g_0^2 n_s^2+1\right) ({\Delta
}-\Delta_{\rm eq}) e^{-\frac{1}{4} g_0 n_s {\Delta_{\rm eq}}}.
\end{equation}
The restoring force ($\sim \Delta$) originates from the phase-dependent terms,
$\cos{(n_d\,n_s\,g_0^2\,t/8)}$. This appealing form of the force of interaction
shows that the force between the solitons is of Hooke's law type with a spring
constant
\begin{equation}
k=\frac{m}{64} g_0^3 n_s^3 \left(g_0^2 n_s^2+1\right) e^{-\frac{1}{4} g_0 n_s
{\Delta_{\rm eq}}},
\end{equation}
where $m$ is the {\it bare mass} of the molecule. Expressed in terms of $n_s$
and $n_d$, the spring constant takes the form
\begin{equation}
k=\frac{g_0^4 n_d^2 n_s^2 \left(g_0^2 n_s^2+1\right){}^2}{64 \left(g_0^2
   n_s^2-1\right)^2}\,m\label{keq},
\end{equation}
which shows that $k=0$ for $n_d=0$; corresponding to a soliton molecule of
infinite bond length. Furthermore, $k$ diverges for $n_s\,g_0=1$, which
signifies soliton coalescence, as we have pointed out in the previous
subsection. Since the frequency of the soliton molecule is given by
\begin{equation}
\omega={1\over8}\,n_d\,n_s\,g_0^2\label{freqeq},
\end{equation}
and the spring constant is given in Eq.~(\ref{keq}), the effective mass
$m^*=k/\omega^2$ will be given by
\begin{equation}
m^*=\frac{\left(g_0^2 n_s^2+1\right){}^2}{\left(g_0^2
n_s^2-1\right)^2}\,m\label{masseq},
\end{equation}
which again diverges at the soliton coalescence condition, $n_s\,g_0=1$. Having
determined the main properties of the soliton molecule, we can now return back
to Eq.~(\ref{delmol}) to express $\Delta$ as
\begin{equation}
{\Delta}={\Delta_0}+\frac{4}{g_0 n_s} \log \left(\frac{g_0^2 n_s^2+2 g_0 n_s
\cos (\omega\,t )+1}{\left(g_0
   n_s+1\right){}^2}\right),
\end{equation}
where $\Delta_0$ is the initial value of $\Delta$, which is given by the
solitons parameters through
\begin{equation}
{\Delta_0}={\Delta_{\rm eq}}+\frac{4}{g_0\,n_s}\, \log \left(\frac{1+g_0^2
{n_s}^2}{\left(1-g_0{n_s}\right)^2}\right).
\end{equation}
The amplitude of the oscillation $\Delta_{\rm max}=\Delta_0-\Delta_{\rm eq}$ is
thus given by
\begin{equation}
\Delta_{\rm max}=\frac{4}{g_0\,n_s}\, \log \left(\frac{1+g_0^2
{n_s}^2}{\left(1-g_0{n_s}\right)^2}\right),
\end{equation}
which gives an elastic potential energy
\begin{equation}
E={1\over2}\,k\,\Delta_{\rm max}^2={1\over8}\,g_0^2\,n_d^2\,{\left(1+g_0^2
{n_s}^2\over {1-g_0^2
   {n_s}^2}\right)^2}\,{\log ^2\left(\frac{1+g_0^2 {n_s}^2}{\left(1-g_0
{n_s}\right)^2}\right)}.
\end{equation}
It should be noted here that this is equal to the mechanical energy since the
initial speed vanishes, ${\dot\Delta}(0)=0$. The fact that the potential energy
diverges at the coalescence condition $n_s\,g_0=1$ is a gain an artifact of the
calculation, but it at least indicates that the bond is tighter than cases
where $n_s\,g_0\gg1$ or $n_s\,g_0\ll1$.

\subsection{Stability}
\label{stability_sec}  Here, we investigate the stability of the soliton
molecule against break up in the following three collision regimes: $i)$
reflection by a hard wall, $ii)$ crossing a finite potential barrier, and
$iii)$ collision with a single soliton. To that end we solve the
Gross-Pitaevskii equation, Eq.~(\ref{gp}), numerically. As an initial state, we
use, for cases $i)$ and $ii)$ the two solitons solution, Eq.~(\ref{exactsol}),
which represents the soliton molecule. For case $iii)$, we use the
superposition of the exact single soliton, Eq.~(\ref{psi1_exact}), with the two
solitons solution.

\begin{figure}
\begin{center}
\includegraphics[width=15cm]{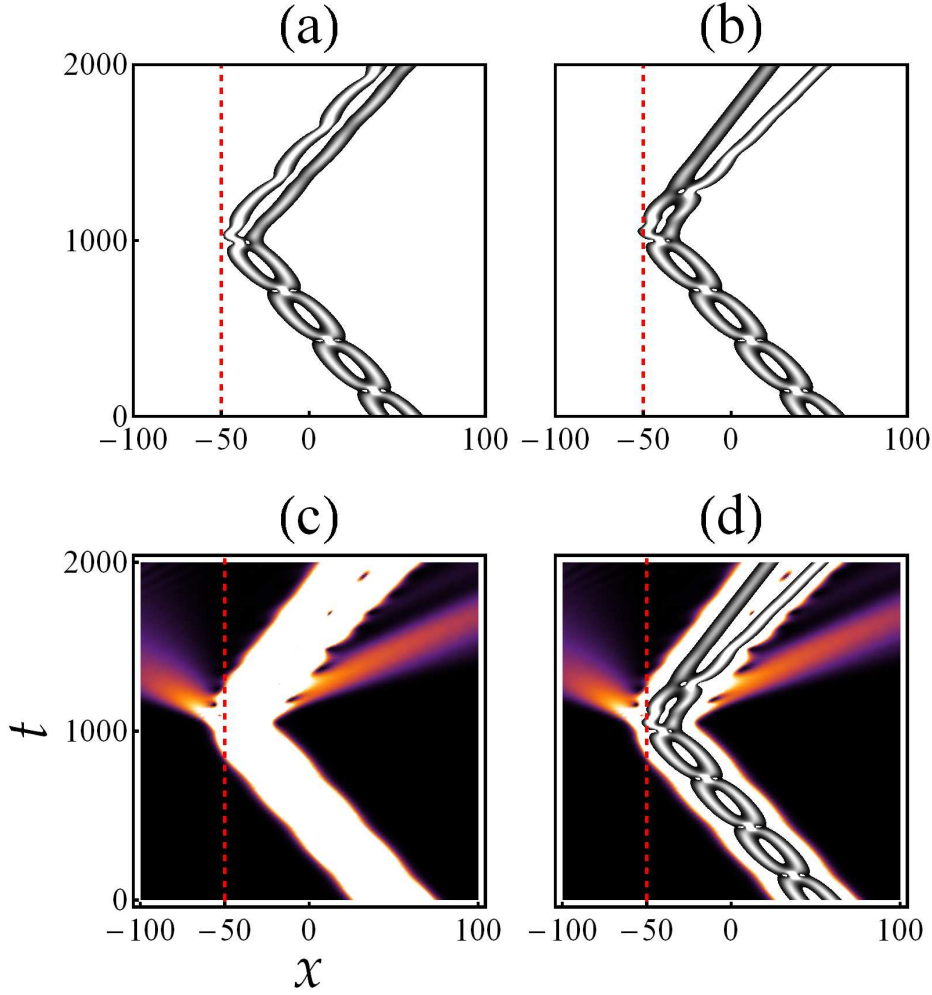}
\end{center}
\caption{Soliton molecule reflection by a potential step given by
Eq.~(\ref{potstep}). The dashed vertical line represents the interface of the
potential step. The parameters used are: $g_0=0.5$, $x_2=x_1=50$,
$\phi_{01}=\phi_{02}=c=0$, $n_d=0.3$, $n_1=1.0$, $\eta=-0.1$, $x_0=-50$. {\bf
(a)} $V_0=100$. {\bf (b),(c),(d)} $V_0=0.075$. } \label{fig9}
\end{figure}

\begin{figure}
\begin{center}
\includegraphics[width=15cm]{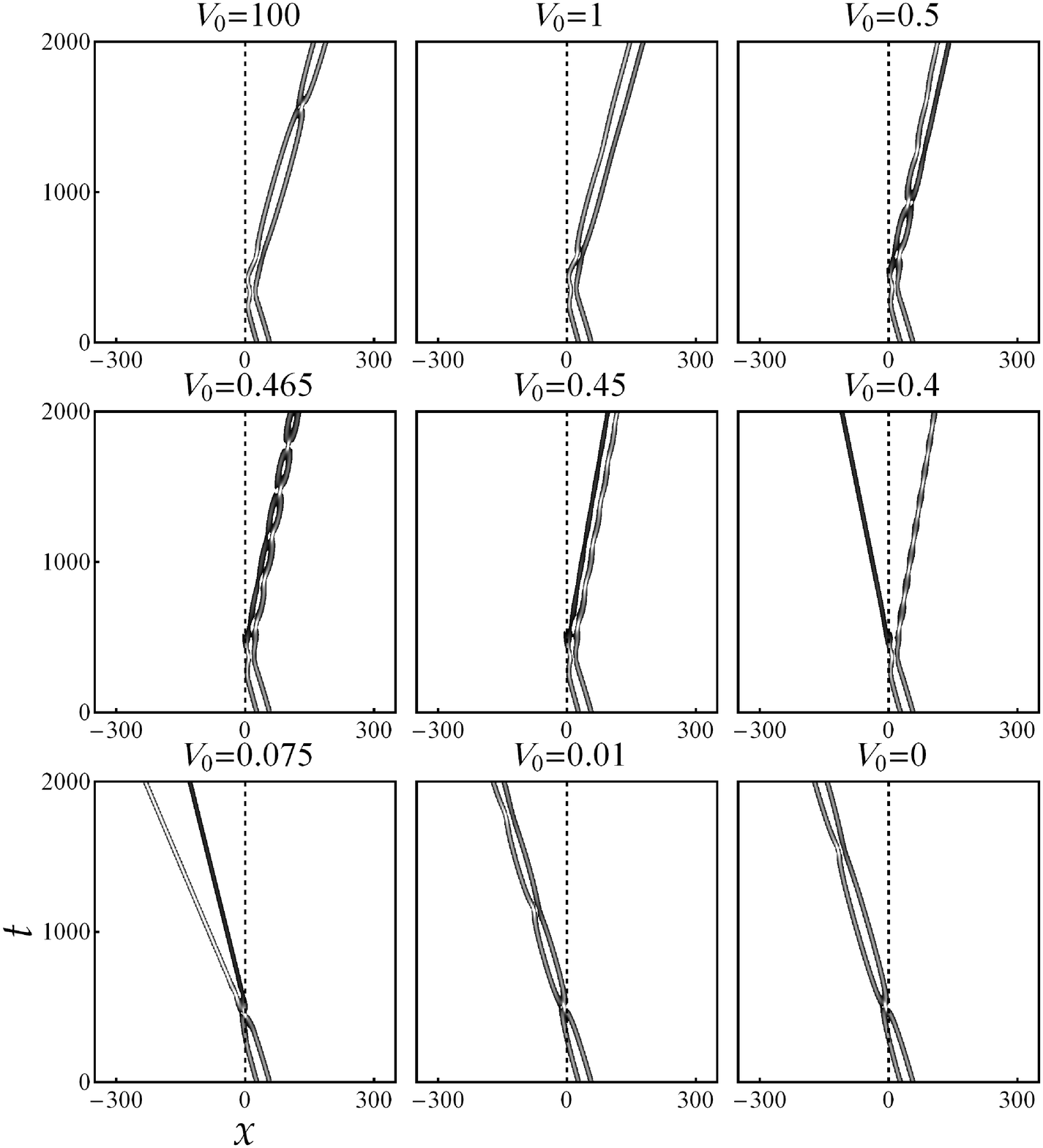}
\end{center}
\caption{Soliton molecule reflection by and transmitting through a potential
barrier given by Eq.~(\ref{potbarr}) for different barrier heights. The
parameters used are: $g_0=0.5$, $x_2=x_1=42$, $\phi_{01}=\phi_{02}=c=0$,
$n_d=0.1$, $n_1=0.95$, $\eta=-0.1$, $d=0.005$, $x_0=0$.} \label{fig10}
\end{figure}

\begin{figure}
\begin{center}
\includegraphics[width=15cm]{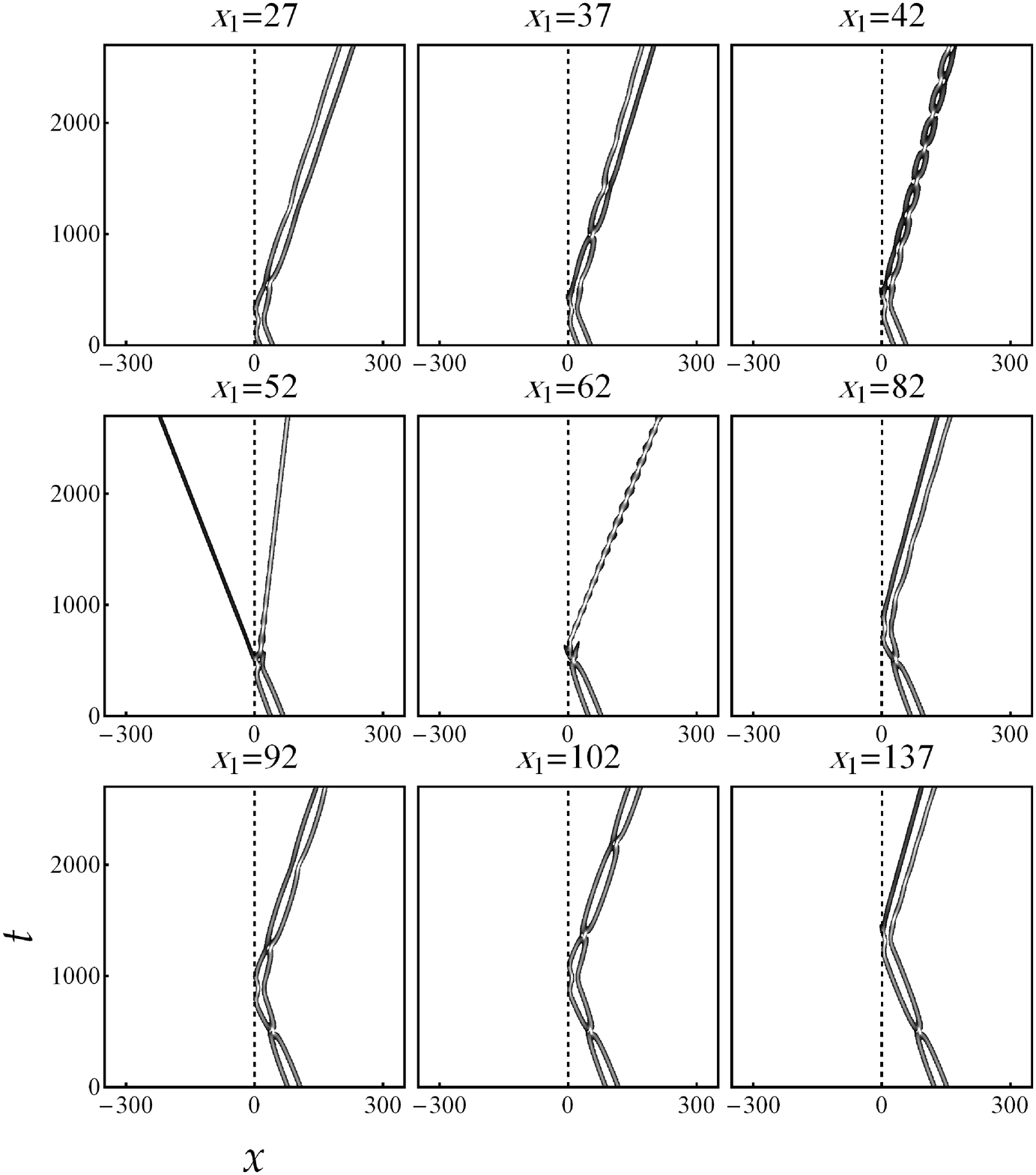}
\end{center}
\caption{Soliton molecule reflection by a potential barrier given by
Eq.~(\ref{potbarr}) for different initial positions of the molecule. The
parameters used are: $x_2=x_1$, $g_0=0.5$, $\phi_{01}=\phi_{02}=c=0$,
$n_d=0.1$, $n_1=0.95$, $\eta=-0.1$, $V_0=0.465$, $d=0.005$, $x_0=0$.}
\label{fig11}
\end{figure}

\begin{figure}
\begin{center}
\includegraphics[width=15cm]{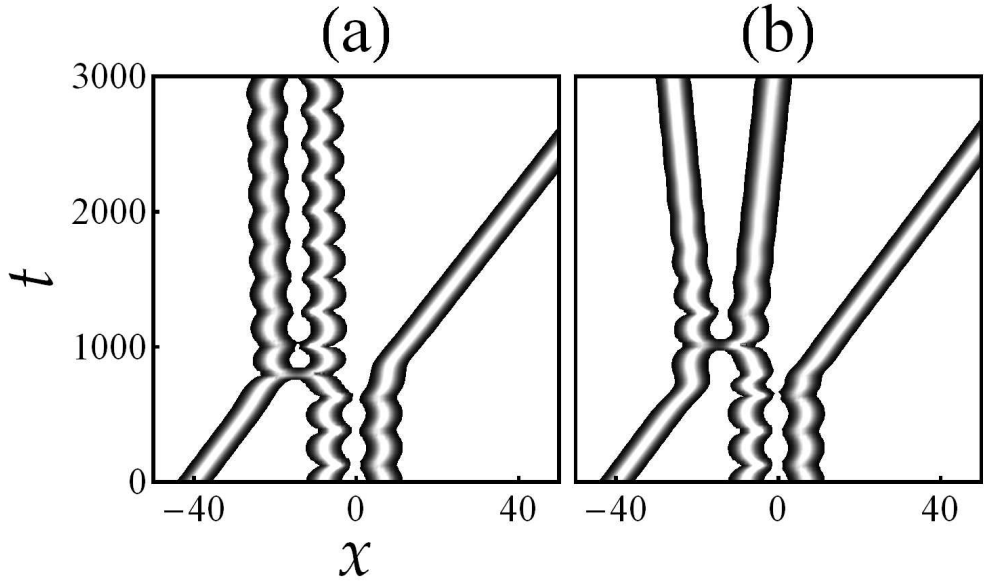}
\end{center}
\caption{Collision between a single soliton and a soliton molecule for two
single soliton phases. $g_0=0.5$, $x_2=x_1=\phi_{01}=\phi_{02}=c=0$, $n_d=0.2$,
$n_1=1.9$, $n_3=2$, $x_3=-40$, $\eta_3=0.025$. The phase difference between the
injected soliton of {\bf (b)} and that of {\bf(a)} equals $\pi$. }
\label{fig12}
\end{figure}

\begin{figure}
\begin{center}
\includegraphics[width=15cm]{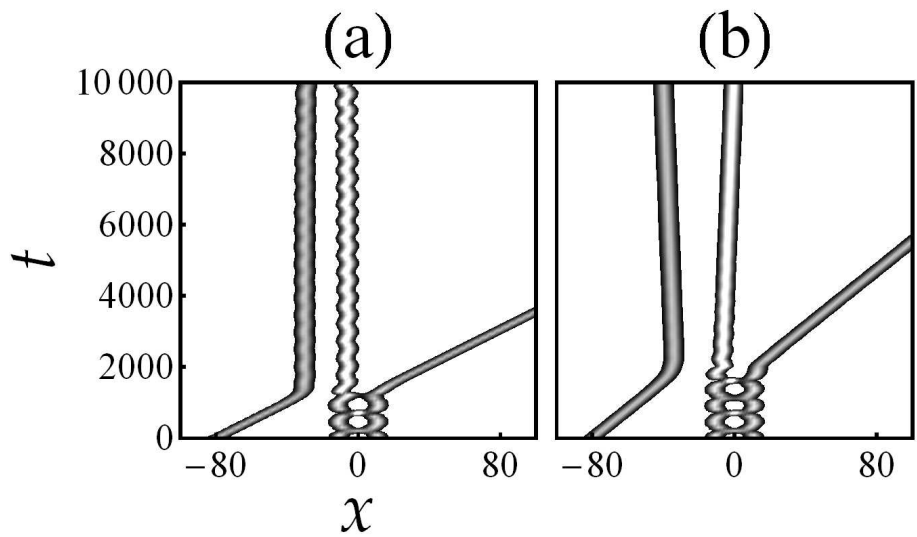}
\end{center}
\caption{Collision between a single soliton and a soliton molecule for two
single soliton initial speeds. $g_0=0.5$, $x_2=x_1=\phi_{01}=\phi_{02}=c=0$,
$n_d=0.2$, $n_1=1$, $n_3=1$, $x_3=-80$. {\bf (a)} $\eta_3=-0.04$, {\bf (b)}
$\eta_3=-0.025$. } \label{fig13}
\end{figure}

\begin{figure}
\begin{center}
\includegraphics[width=15cm]{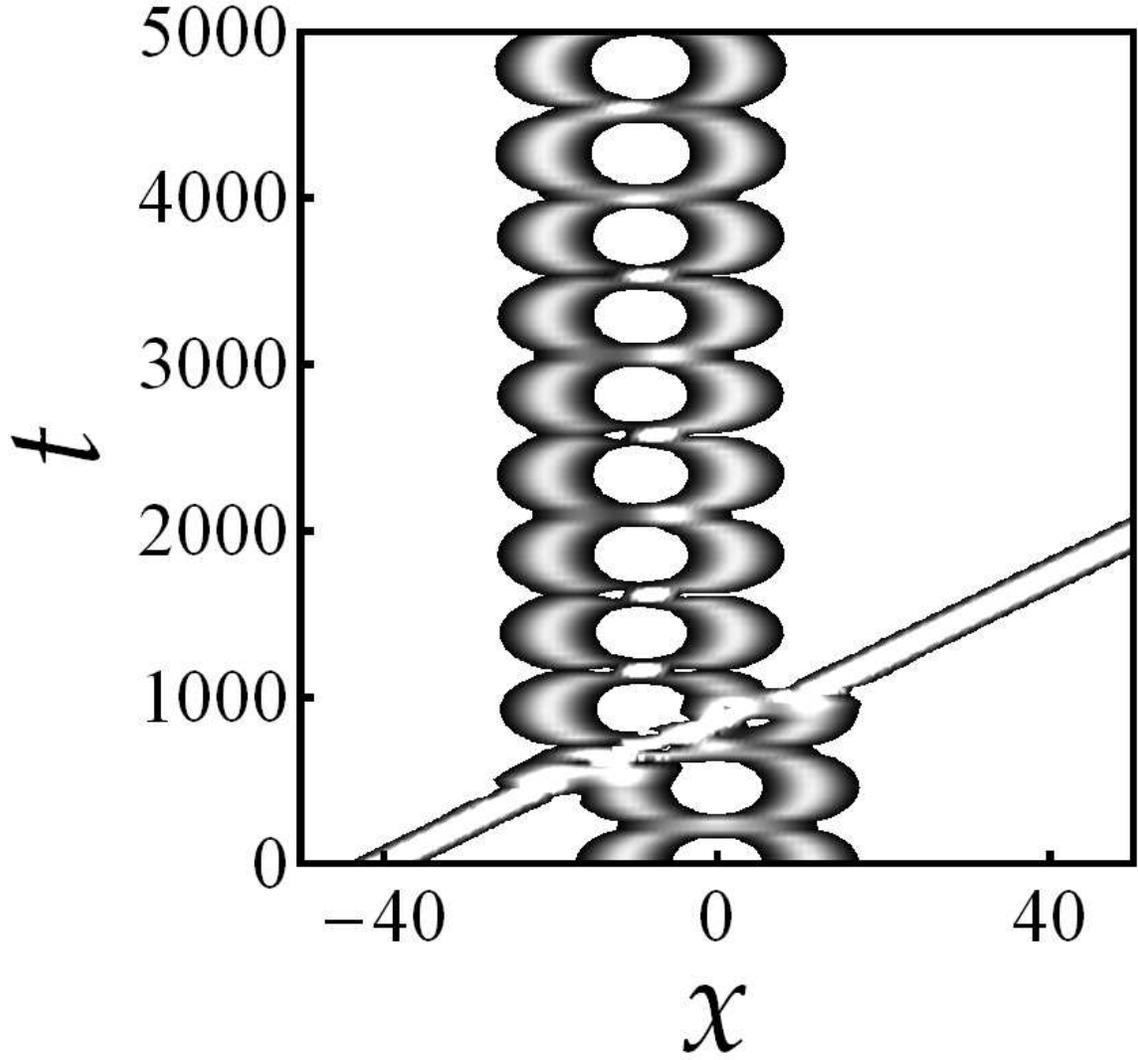}
\end{center}
\caption{Collision between a high intensity single soliton and a soliton
molecule. $g_0=0.5$, $x_2=x_1=\phi_{01}=\phi_{02}=c=0$, $n_d=0.2$, $n_1=1$,
$n_3=2$, $\eta_3=0.04$, $x_3=-40$. } \label{fig14}
\end{figure}

Before starting the discussion of results, we point out that in
Figs.~\ref{fig9}-\ref{fig14}, we present the results of this section using
spacio-temporal density plots. Since the solitons are too thin compared to the
spacial range that we consider, a density plot with full spacial and time
ranges will not show a clear solitons peak density or center-of-mass path, as
Fig.~\ref{fig9}c shows. To solve this problem, we restricted the density
plotting to a finite range of $|\Psi(x,t)|^2$, namely between 0.025 and 0.15
corresponding to the upper part of the solitons peaks. This results in an
easier tracking of both the solitons peak density and center-of-mass path, as
shown in Fig.~\ref{fig9}a,b,d and the rest of subsequent figures.

For reflection from a hard wall we solve Eq.~(\ref{gp}) with a potential step
of the form
\begin{equation}
V(x)=\left\{
\begin{array}{cc}
V_0\,,\hspace{1cm}&x<x_0,\\\\
0\,,\hspace{1cm}&x\geq x_0,\end{array} \right.\label{potstep}
\end{equation}
where $V_0$ and $x_0$ are the hight and location of the potential wall,
respectively. The result of reflection from this hard wall, with $V_0=100$, is
shown in Fig.~\ref{fig9}a. The soliton molecule preserves its molecular
structure but with different characteristics. The solitons in the reflected
molecule do not coalesce as in the incident molecule. In other words, the
equilibrium bond length becomes larger. The density plot shows that initially
the two solitons are of comparable intensities. After reflection, the brighter
color of the left soliton and darker color of the right soliton indicate that
the left soliton acquires higher intensity on the expense of the right soliton.
We also notice that the left soliton performs two reflections from the
potential interface. After the first reflection, it collides with the right
soliton and then collides with the potential interface for the second time.

The picture becomes different when the hight of the wall is reduced to
$V_0=0.075$, as shown in Fig.~\ref{fig9}b. The soliton molecule breaks up after
reflection. This is due to loss of energy at the interface of the potential.
Part of the soliton molecule transmits as a nonsolitonic  pulse that broadens
and decays in intensity by time. By plotting $|\Psi(x,t)|^2$ in
Fig.~\ref{fig9}c with its full range, we can see the nonsolitonic part as the
left- and right-going two red ejections corresponding to the transmitted and
reflected nonsolitonic pulses, respectively. In Fig.~\ref{fig9}d, we combine
Fig.~\ref{fig9}b and Fig.~\ref{fig9}c to show the locations of the nonsolitonic
ejections with respect to the solitons centers. In the case of reflection from
a hard wall, the nonsolitonic ejections are essentially not present which
results in the stability of the molecular structure.

For reflection from a potential barrier we solve Eq.~(\ref{gp}) with the
potential
\begin{equation}
V(x)=\left\{
\begin{array}{cc}
V_0,\hspace{1cm}&x_0-d<x<x_0,\\\\
0\,\hspace{1cm}&{\rm elsewhere},
\end{array} \right.
\label{potbarr}
\end{equation}
where $V_0$, $d$, and $x_0$, are the height, width, and location of the right
side of the barrier, respectively. In Fig.~\ref{fig10}, we show the many
different possibilities that result when the height of the barrier is changed.
The free evolution case with $V_0=0$ is shown as a reference plot. The full
reflection case is shown for $V_0=100$, which is similar to the previous case
of reflection from a hard wall. Reducing the height of the barrier to $V_0=1$,
we notice that the soliton molecule breaks up after reflection. As pointed
above, this is due to the nonsolitonic ejections taking place at the interfaces
of the potential. Reducing the height of the barrier to $V_0=0.5$, a sign of
solitons recombining appears in the form of a soliton molecule of a short
lifetime. At $V_0=0.465$ a stable molecule is remarkably formed with a
considerably shorter period than for the incident molecule. We have confirmed
numerically that this molecule remains stable for much longer time provided
that the soliton molecule remains sufficiently far from the boundaries of the
spacial grid. This unique structure remains for some small domain around
$V_0=0.465$, but is lost for $V_0=0.45$, where the molecule breaks for long
evolution times. Decreasing the height of the barrier to $V_0=0.4$, the
molecule breaks at the interface and splits into a reflected and transmitted
solitons. For $V_0=0.075$, the molecule breaks at the interface, but both
solitons transmit through the barrier. For $V_0=0.01$, the transmitted solitons
show a sign of recombining again but with shorter period than for the free
evolution case and larger than in the case of $V_0=0.465$.

Motivated by the fact that at the coalescence point the intensity of the
molecule is considerably higher than at other instants, we expect to find
different scattering dynamics when the soliton molecule meets the interface of
the potential at different phases of its periodic oscillation. In
Fig.~\ref{fig11}, this is investigated by fixing the height of the potential
barrier at $V_0=0.465$ and changing the initial launching position of the
molecule. Starting at $x_1=27$, the molecule breaks after reflection. A
shortly-lived molecule is obtained at $x_1=37$, and a stable molecule is found
for $x_1=42$, which corresponds to the $V_0=0.465$ case of Fig.~\ref{fig10}. At
$x_1=52$, the soliton splits at the interface into transmitted and reflected
solitons. The coalescence point is, in this case, located at the interface.
Transmission takes place due to the high intensity of the soliton at the
coalescence point. At $x_1=62$, the two solitons still split as in the previous
subfigure but with a weak transmitted soliton intensity less than 0.025 and
hence will not be shown in our density plots which are restricted to the
intensities between 0.025 and 0.15, as pointed out previously. For $x_1=82$,
the coalescence point takes place before the molecule reaches the interface and
both solitons reflect but the molecule breaks up. For $x_1=92$ the two
reflected solitons start to recombine forming a stable molecule at $x_1=102$.
At $x_1=137$ the reflected molecule starts to break up since the second
coalescence point becomes close to the interface. Thus, the conclusion from
this figure is that the soliton molecule is more vulnerable to break up when it
meets the interface at the coalescence point. Equivalently, soliton molecules
with larger equilibrium bond length, such that coalescence does not occur, will
be more stable against breakup post reflection from barriers.

Finally, we present in Figs.~\ref{fig12}-\ref{fig14} the results of scattering
of a soliton molecule by a single soliton described by Eq.~(\ref{newsol1}) with
normalization $n_3$, center-of-mass position and speed $x_3$ and $\eta_3=v_3$,
respectively. The effects of the phase, speed, and amplitude of the injected
soliton are investigated separately. In Fig.~\ref{fig12}a, a soliton initially
at $x=-40$ is launched towards a stationary soliton molecule near $x=0$. At the
impact, the molecule brakes up, its right soliton is ejected in the direction
of the positive $x$-axis, and the left soliton combines with the scatterer
soliton to form a new stationary molecule shifted by about a bond length to the
left. We point out here that for such an outcome to occur, it is essential that
the amplitude of the scatterer soliton is nearly equal to that of the right
soliton of the molecule. Otherwise, a different outcome, as that of
Fig.~\ref{fig14}, will be obtained. In Fig.~\ref{fig12}b, the same numerical
experiment is repeated but with adding a $\pi$ to the phase of the scatterer
soliton. Clearly, this phase addition prevents the formation of a new molecule
resulting in three solitons diverging from each other. From applications point
of view, the phase of $\pi$ could be used as a ``key'' to ``unlock'' the
molecule for the purpose of extracting stored data. In Fig.~\ref{fig13}, we
show the effect of the initial speed of the injected soliton. In contrary to
one's first judgment, the molecule preserves its structure for fast collisions,
as in Fig.~\ref{fig13}a, and breaks up for slower collisions, as in
Fig.~\ref{fig13}b. In Fig.~\ref{fig14}, the injected soliton has an amplitude
that is approximately two times larger than any of the two solitons of the
molecule. The injected soliton penetrates the molecule leaving it almost
unchanged apart from a center-of-mass shift to the left.

\section{Conclusions}
\label{conc_sec} We have used the Inverse Scattering method to derive the two
solitons solution of a nonlinear Schr$\rm\ddot o$diner equation with a
parabolic potential and cubic nonlinearity with time-dependent coefficients, as
given by Eq.~(\ref{gp}). The solution was then simplified and put in a
suggestive form in terms of the fundamental parameters of the two solitons,
namely their amplitudes, center-of-mass positions and speeds, and their phases.
In this form, two different regimes of the solution, namely the resolved
solitons and overlapping solitons, were distinguished and the main features
such as the solitons separation and relative phase were extracted. From the
expression for the solitons separation we find that for the homogeneous case
and zero solitons relative speed, the solitons separation diverges
logarithmically with the solitons amplitude difference such that, for equal
solitons, the trivial solution is obtained.

The force of interaction was then derived, essentially, for arbitrary solitons
parameters. This resulted in generalizing Gordon's formula \cite{gordon} to
{\it i}) the generalized inhomogeneous case considered here, {\it ii})
arbitrary solitons relative speed and amplitudes, and {\it iii}) short solitons
separations (compared to their width). With this formula, the possibility of
forming soliton molecules emerged naturally, where the force at short distance
was shown to be composed of an attractive part, resulting from the
nonlinearity, and another part that oscillates between repulsive and attractive
resulting from the time-dependent relative phase. The main features of the
soliton molecule, including its equilibrium bond length and bond spring
constant, frequency and amplitude of oscillation, effective mass, and its
elastic potential energy, where then calculated in terms of the solitons
parameters. It turns out that the amplitudes difference $n_d=n_2-n_1$ plays an
important role in determining these quantities. Furthermore, we show that at
the condition $g_0\,n_s=1$, the two solitons coalesce while away from this
condition, the solitons approach each other but remain resolved. At this
condition, the molecule's effective mass and spring constant have maximum
values. In our expressions Eqs.~(\ref{keq}) and (\ref{masseq}) diverge because
these formulae were derived assuming the solitons remain resolved.

To have a sense of its stability we investigated numerically {\it i}) the
collision of the soliton molecule with a hard wall and softer one, {\it ii})
scattering by a potential barrier, and {\it iii}) collision with a single
molecule. The first case showed that while the molecular structure is preserved
after reflection from a hard wall, it breaks when reflecting from a softer one.
Reflection from a finite barrier showed that the molecular structure is
preserved only for specific heights of the barrier. For an incident molecule
with the coalescence condition satisfied, the molecule will be most vulnerable
to break up when the coalescence point takes place at the interface of the
barrier. This is simply understood by the fact that the intensity of the
soliton molecule is maximum when the two solitons coalesce such that tunneling
becomes possible. Stability of the molecule was also investigated by scattering
the molecule with a single soliton. It turned that slower collisions tend to
break up the molecule more easily than faster ones. In addition, the outcome of
the collision depends on the phase of the incoming soliton such that a
scatterer soliton which is in-phase with the molecule will typically preserve
its molecular structure, but for an out-of-phase soliton, the molecule breaks
up.

The two solitons solution presented here and the analysis that shows how to
extract the solitons separation and relative phase may constitute the basis for
a more accurate and detailed investigation of the origin of the soliton-soliton
force, especially for short separations and at coalescence. The results of this
paper will be hopefully of relevance to possible future applications of soliton
molecules as data carriers or memories.

\section*{Acknowledgement} The author acknowledges helpful discussions with
Vladimir N. Serkin.

\appendix
\section{Deriving the two solitons solution}
\label{appa} The Lax pair associated with the Gross-Pitaevskii equation,
Eq.~(\ref{gp}), is obtained using our Lax pair search method
\cite{usama_lax,usama_vib} and reads
\begin{equation}
{\bf\Phi}_x=\zeta\,{\bf J}\cdot{\bf\Phi}\cdot{\bf\Lambda}+{\bf P}\cdot{\bf\Phi}
\label{phix},
\end{equation}
\begin{equation}
{\bf\Phi}_t=i\,\zeta^2\,{\bf
J}\cdot{\bf\Phi}\cdot{\bf\Lambda}\cdot{\bf\Lambda}+\zeta\,\left(i\,{\bf
P}+x\,{\dot\gamma}\,{\bf J}\right)\cdot{\bf\Phi}\cdot{\bf\Lambda}+{\bf
W}\cdot{\bf\Phi} \label{phit},
\end{equation}
where\\\\
\begin{math}
{\bf\Phi}(x,t)=\left(\begin{array}{cc}\psi_1(x,t)&\psi_2(x,t)\\\phi_1(x,t)&\phi_2(x,t)\end{array}\right)
\end{math},
\hspace{0.25cm}
\begin{math}
{\bf\Lambda}=\left(\begin{array}{cc}\lambda_1&0\\0&\lambda_2\end{array}\right)
\end{math},
\hspace{0.25cm}
\begin{math}
{\bf J}=\left(\begin{array}{cc}1&0\\0&-1\end{array}\right)
\end{math},
\hspace{0.25cm}
\begin{math}
{\bf
P}=\left(\begin{array}{cc}0&\sqrt{g_0}\,Q\\-\sqrt{g_0}\,Q^*&0\end{array}\right)
\end{math},\\\\\\
\begin{math}
{\bf W}=\left(\begin{array}{cc} i\,g_0\,|Q|^2/2&
\sqrt{g_0}\,x\,{\dot\gamma}\,Q+i\,\sqrt{g_0}\,Q_x/2\\
-\sqrt{g_0}\,x\,{\dot\gamma}\,Q^*+i\,\sqrt{g_0}\,Q^*_x/2& -i\,g_0\,|Q|^2/2
\end{array}\right)
\end{math},\\\\\\
$\zeta(t)=\sqrt{2}\,e^{\gamma(t)}$,
$Q(x,t)=\Psi(x,t)\,e^{(\gamma(t)+i\,{\dot\gamma}(t)\,x^2)/2}$, and $\lambda_1$
and $\lambda_2$ are arbitrary constants. Here, $\Psi(x,t)$ is the solution of
Eq.~(\ref{gp}) and ${\bf\Phi}$ is the {\it auxiliary} field.

The {\it compatibility condition} ${\bf\Phi}_{xt}={\bf\Phi}_{tx}$ of the linear
system, Eqs.~(\ref{phix}) and (\ref{phit}), is equivalent to the
Gross-Pitaevskii equation, Eq.~(\ref{gp}), and its complex conjugate. For a
known {\it seed} solution, $\Psi_0(x,t)$, of Eq.~(\ref{gp}) the linear system
will have the solution ${\bf\Phi_0}$. The Darboux transformation is defined as
${\bf\Phi}[1]={\bf\Phi}\cdot{\Lambda}-{\bf\sigma}\,{\bf\Phi}$, where
${\bf\Phi}[1]$ is the transformed field and
${\bf\sigma}={\bf\Phi_0}\cdot{\bf\Lambda}\cdot{\bf\Phi_0}^{-1}$. Requiring the
linear system to be {\it covariant} under the Darboux transformation imposes
the transformation ${\bf P}[1]={\bf P}+{\bf
J}\cdot{\bf\sigma}-{\bf\sigma}\cdot{\bf J}$, where ${\bf P}[1]$ is the
transformed ${\bf P}$. This gives the new solution in terms of the seed
solution as
\begin{equation}
\Psi(x,t)=\Psi_0(x,t)-\sqrt{8\over
g_0}\,(\lambda_1-\lambda_2)\,e^{(\gamma(t)-i\,{\dot\gamma}(t)\,x^2)/2}
{\psi_1(x,t)\,\psi_2(x,t)\over\phi_2(x,t)\,\psi_1(x,t)-\phi_1(x,t)\,\psi_2(x,t)}
\label{newsol1}.
\end{equation}

Using the trivial solution $\Psi(x,t)=0$ as a seed, the Darboux transformation
generates the well-known sech-shaped single soliton solution \cite{usama_vib}
\begin{equation}
\psi(x,t)=\sqrt{n_1\,\alpha_{11}\over2}\,e^{i\phi_1(x,t)}\,{\rm
sech}\left({\alpha_{11}(x-x_{\rm cm1}(t))}\right) \label{psi1_exact},
\end{equation}
where
\begin{equation}
\phi_1(x,t)=\phi_0(t) +{\dot x_{\rm cm1}}(t)\,(x-x_{\rm
cm1}(t))-{1\over2}{\dot\gamma}(t)(x-x_{\rm cm1}(t))^2 \label{pha},
\end{equation}
\begin{equation}
x_{\rm cm1}(t)=\left(x_1\,e^{2\gamma_0}+\eta_1\,
g(t)\right)e^{-\gamma(t)-\gamma_0} \label{xcm1},
\end{equation}
\begin{equation}
\phi_0(t)=\phi_{01}+{1\over8}\,g(t)\,\left(4\,e^{-2\gamma_0}\eta_1^2+(n_1g_0)^2\right)-{1\over2}\,{\dot\gamma}(t)\,x_{\rm
cm1}^2(t)
\end{equation}
$g(t)=\int_{0}^t e^{2\gamma(t^\prime)}dt^\prime$,
$\alpha_{11}=e^{\gamma(t)}\,n_1\,g_0/2$, $\gamma(0)=\gamma_0$, and the constant
$\phi_{01}$ corresponds to an arbitrary overall phase. This solution
corresponds to a soliton density profile that is localized at $x_{\rm cm1}(t)$,
moving with center-of-mass speed ${\dot x}_{\rm cm1}(t)$, and normalized to
$n_1$.

Using this single soliton solution as a seed, the Darboux transformation
generates a two solitons solution. The solution of the linear system,
Eqs.~(\ref{phix}) and (\ref{phit}), can in this case be derived and simplified
to the following form
\begin{equation}
{\psi_1}(x,t)=e^{\frac{\sqrt{2} y {\lambda_1}}{g_0 {n_1}}+{y_2}}
\left(\frac{\exp \left({y_1}+y
   \left(\frac{1}{2}+\frac{-2 \sqrt{2} {\lambda_1}+i e^{-\gamma_0} {\eta_1}}{g_0
   {n_1}}\right)\right)}{e^y+1}-\frac{g_0 {n_1} \left(e^y-1\right)}{e^y+1}+2 \left(2 \sqrt{2}
    {\lambda_1}-i
   e^{-\gamma_0} {\eta_1}\right)\right),
\end{equation}
\begin{equation}
{\psi_2}(x,t)=e^{\frac{-\sqrt{2} y {\lambda_1^*}}{g_0 {n_1}}+{y_3}}
\left(-\frac{2 g_0
   {n_1} \exp \left(-{y_1^*}+y \left(\frac{1}{2}+\frac{2 \sqrt{2} {\lambda_1^*}+i e^{-\gamma
   (0)} {\eta_1}}{g_0 {n_1}}\right)\right)}{e^y+1}+\frac{-2 \sqrt{2} {\lambda_1^*}-i
   e^{-\gamma_0} {\eta_1}}{g_0 {n_1}}-\frac{e^y-1}{2
   \left(e^y+1\right)}\right),
\end{equation}
\begin{equation}
{\varphi_1}(x,t)=\psi_2^*(x,t),
\end{equation}
\begin{equation}
{\varphi_2}(x,t)=-\psi_1^*(x,t),
\end{equation}
\begin{equation}
\lambda_2=-\lambda_1^*,
\end{equation}
where,
\begin{equation}
y=g_0 {n_1} e^{\gamma (t)} (x-x_{\rm cm1}(t)),
\end{equation}
\begin{equation}
y_1=\frac{1}{8} e^{-2 \gamma_0} \left(-i e^{2 \gamma_0} g(t) \left(32
{\lambda_1}^2-g_0^2
   {n_1}^2\right)-16 \sqrt{2} e^{\gamma_0} {\eta_1} {\lambda_1} g(t)+4 i
   {\eta_1}^2 g(t)\right),
\end{equation}
\begin{equation}
y_2 = \left(\sqrt{2} e^{-\gamma_0} \eta_1 + 2 i \lambda_1\right) \lambda_1
g(t),
\end{equation}
\begin{equation}
y_3=-{\lambda_1^*} g(t) \left(\sqrt{2} e^{-\gamma_0} {\eta_1}-2 i
{\lambda_1^*}\right)+i {\varphi_{01}}.
\end{equation}

Finally, the two solitons solution is obtained by substituting for $\psi_{1,2}$
and $\phi_{1,2}$ in Eq.~(\ref{newsol1}), which upon substituting
$\lambda_1=n_2\,g_0/(4\sqrt{2})+i\,e^{-\gamma_0}\eta_2/\sqrt{8}$ and further
simplification can then be put in the form of Eq.~(\ref{exactsol}).

\section{Deriving center-of-mass positions and acceleration}
\subsection{Center-Of-Mass Positions}
\label{appb1} Since $X$ and $Y$ are functions of $x$, we start by examining
their values near the roots of $q$. This task can be simplified by rewriting
$X$ as $X=e^{\frac{1}{2}e^{\gamma (t)} g_0(n_1(x-x_{\rm cm1}(t))-n_2(x-x_{\rm
cm2}(t))}$. For $x\simeq x_{\rm cm1}(t)$, we get $X\simeq
e^{\frac{1}{2}e^{\gamma (t)} g_0\,n_2(x_{\rm cm2}(t)-x_{\rm cm1}(t))}\gg1$, and
for $x\simeq x_{\rm cm2}(t)$, we have $X\simeq e^{\frac{1}{2}e^{\gamma (t)}
g_0\,n_1(x_{\rm cm2}(t)-x_{\rm cm1}(t))}\gg1$. For $x>x_{\rm cm2}(t)$  the
first term in the exponent of $X$ is larger than the second one, provided that
$n_1$ is not much larger than $n_2$, as remarked at the end of this section,
hence $X\gg1$. For $x<x_{\rm cm1}(t)$ the magnitude of the first term in the
exponent of $X$ is smaller than the magnitude of the second one, which again
leads to $X\gg1$. For the region $x_{\rm cm1}(t)<x<x_{\rm cm2}(t)$, the
condition $X\gg1$ is always satisfied. In conclusion, $X\gg1$ for all $x$,
apart from situations with extreme values of $n_1/n_2$. The situation is
simpler for $Y$: $Y\gg1$ for $x>x_{\rm cm2}(t)$, $Y\simeq1$ for $x\simeq x_{\rm
cm1}(t)$, and $Y\simeq0$ for $x<x_{\rm cm1}(t)$. Thus, to find $x_r$, we expand
$q$ in powers of large $X$ and $Y$ and to find $x_l$ and $x_c$, we expand $q$
in powers of large $X$.

Expanding $q$ in powers of $X$ and $Y$ around $\infty$ up to first order in
$1/X$ and $1/Y$, we find
\begin{eqnarray}
q&=&\frac{1}{2} \left(\log{Y }-2 \log{X}+\log{Y_r}\right)-\frac{g_0^2
e^{\gamma_0} ({n_d}-{n_s})^2
   \left(g_0 {n_s} e^{\gamma
   (0)} \cos{z}-2 {\eta_d} \sin{z}\right)}
   {X \left(g_0^2 {n_s}^2 e^{2 \gamma_0}+4 {\eta_d}^2\right)}
   \label{q1},
\end{eqnarray}
where
\begin{equation}
Y_r=\frac{g_0^2 ({n_d}-{n_s})^2 \left(g_0^2 {n_d}^2 e^{2 \gamma_0}+4
{\eta_d}^2\right)}{g_0^2 {n_s}^2
   e^{2 \gamma_0}+4 {\eta_d}^2}.
\end{equation}
The root of this equation gives the center-of-mass position of the right
soliton. The first two log terms equal $n_2\,g_0\,(x-x_{\rm cm2}(t))$. The
third log term is constant and diverges for $n_d=\eta_d=0$. The last term is
small since it is proportional to $1/X$, but it is needed because it contains
the phase-dependent contributions. Due to the combination of $\log{X}$ and $X$,
finding an algebraic expression for the root of this equation will not be
possible. Instead, we ignore at first the $1/X$ term to obtain the dominant
contribution which will then be used to find the $1/X$ contribution. This gives
\begin{equation}
x_r=x_{\rm cm2}(t)- {2 e^{-\gamma (t)} \over{g_0
   ({n_d}+{n_s})}}{\log Y_r}.
\end{equation}
Substituting back in Eq.~(\ref{q1}) and solving for $x$, we finally get
\begin{equation}
x_r=x_{\rm cm2}(t)-\frac{2 e^{-\gamma (t)}
   \log{Y_r}}{ g_0 ({n_d}+{n_s})}
   +\frac{4 {Y_r}^{\frac{{n_s}}{{n_d}+{n_s}}}
 \left(g_0 {n_s}
   e^{\gamma_0} \cos{z_r}-2 {\eta_d} \sin{z_r}\right)}{g_0 ({n_d}+{n_s})
   \left(g_0^2 {n_d}^2 e^{2 \gamma_0}
   +4 {\eta_d}^2\right)}\,e^{\frac{1}{4} g_0 {xd}
   ({n_d}-{n_s}) e^{\gamma (t)}-\gamma (t)+\gamma_0}\label{xreq},
\end{equation}
where $z_r=z(x=x_r)$.

For the left and central solitons, we expand $q$ in powers of $X$; leaving $Y$
arbitrary. In this manner, we account for the two roots, $x_l$ and $x_c$,
simultaneously. Expanding $q$ in powers of large $X$, we get
\begin{equation}
q={1\over2}\log {{Y}}+\frac{1}{2}\log \left(\frac{({n_d}-{n_s})^2}{{4 ({Y}
+1)^2 e^{-2 \gamma_0} {\eta_d}^2}/{g_0^2}+({n_d}+{n_s} {Y}
   )^2}\right)+\frac{C}{X}\label{q2},
\end{equation} where
\begin{eqnarray}
C&=&({Y} +1)^2 e^{-\gamma_0} \left[2 g_0^2 e^{2 \gamma_0} {\eta_d} \sin{z}
\left({n_d}^2+{n_s}^2 {Y} \right)+4 g_0 e^{\gamma_0}
   {\eta_d}^2 \cos{z} ({n_d} {Y} +{n_s})\right.\nonumber\\&+&\left. g_0^3 {n_d} {n_s}
   e^{3 \gamma_0} \cos{z} ({n_d}+{n_s} {Y} )+8 ({Y} +1)
   {\eta_d}^3 \sin{z}\right]\nonumber\\&/&{\left(g_0^2 e^{2 \gamma_0}
   ({n_d}+{n_s} {Y} )^2+4 ({Y} +1)^2 {\eta_d}^2\right)}.
\end{eqnarray}
Similar to the above procedure for $x_r$, we solve first Eq.~(\ref{q2}) without
the $1/X$ term, which gives
\begin{equation}
Y_\pm =\frac{g_0 e^{\gamma_0} \left[g_0 e^{\gamma_0} \left({n_d}^2-4 {n_d}
{n_s}+{n_s}^2\right)\pm\sqrt{M} ({n_d}-{n_s})\right]-8
   {\eta_d}^2}{2 g_0^2 {n_s}^2 e^{2 \gamma_0}+8 {\eta_d}^2},
\end{equation}
where
\begin{equation}
M=g_0^2 e^{2 \gamma_0} \left({n_d}^2-6 {n_d} {n_s}+{n_s}^2\right)-32
{\eta_d}^2.
\end{equation}
In the limit $n_d\rightarrow0$ and $\eta_d\rightarrow0$, the solution $Y_+$
approaches 0, which corresponds to $x_l=x_{\rm
cm1}(t)+(\log{Y_+})/(n_1\,g_0\,e^{\gamma(t)})\rightarrow-\infty$. The solution
$Y_-$ approaches 1, which corresponds to $x_c=x_{\rm cm1}(t)$. Thus, the
solutions $Y_+$ and $Y_-$ correspond the left and central solitons,
respectively. Since we will be interested only in the left and right solitons,
we take for the rest of this section the $Y_+$. To find the contribution of the
$1/X$ term, we substitute for $Y_+$ in the $1/X$ term of Eq.~(\ref{q2}), and
then solve for $Y$ to get a corrected expression for $Y_+$
\begin{equation}
{Y_{+}}=\frac{g_0 e^{\gamma_0} \left[({n_d}-{n_s}) \sqrt{M_+}+g_0
   e^{\gamma_0} \left(e^{\frac{2 {C_+}}{{X_+}}}
   ({n_d}-{n_s})^2-2 {n_d} {n_s}\right)\right]
   -8 {\eta_d}^2}{2 g_0^2 {n_s}^2 e^{2
   \gamma_0}+8 {\eta_d}^2},
\end{equation}
where $M_+=g_0^2 e^{2 \left(\frac{{C_+}}{{X_+}}+\gamma_0\right)}
\left(e^{\frac{2
   {C_\pm}}{{X_+}}} ({n_d}-{n_s})^2-4 {n_d}
   {n_s}\right)-16 {\eta_d}^2 \left(e^{\frac{2
   {C_\pm}}{{X_+}}}+1\right)$, ${X_+}=e^{\frac{1}{4} g_0 {x_d} ({n_d}+{n_s})
   e^{\gamma (t)}+\gamma_0}$,
   $x_d=x_{\rm cm2}(t)-x_{\rm cm1}(t)$, $C_+=C(Y=Y_+, x=x_l, z=z_l)$,
   and $z_l=z(x_l)$.
   Expanding for
large $X_+$ and then solving for $x$, we finally get
\begin{equation}
x_l={x_{\rm cm1}}(t)-\frac{2 e^{-\gamma (t)} \log ({Y_+})}
   {g_0 ({n_d}-{n_s})}-\frac{4 {C_+} \,e^{\gamma_0-\gamma (t)}}{{X_+}
\sqrt{g_0^2 e^{2 \gamma_0} \left({n_d}^2-6 {n_d} {n_s}+{n_s}^2\right)-32
{\eta_d}^2}}\label{xleq}.
\end{equation}

Final remarks about the validity of the above derivation are in order. The
condition $X\gg1$ will be met in the region $x>x_{\rm cm2}(t)$ only for
$n_1/n_2>(x_r-x_{\rm cm2}(t))/(x_r-x_{\rm cm1}(t))$. Note that the numerator of
the right hand side of this inequality is less than the denominator by at least
$x_{\rm cm2}(t)-x_{\rm cm1}(t)$. For $x<x_{\rm cm1}(t)$, the condition $X\gg1$
will be met only for $n_1/n_2<(x_{\rm cm2}(t)-x_l)/(x_{\rm cm1}(t)-x_l)$. Here,
the numerator of the right hand side of this inequality is larger than the
denominator by at least $x_{\rm cm2}(t)-x_{\rm cm1}(t)$. A more quantitative
estimate for the ratio $n_1/n_2$ can be obtained using the above results for
$x_l(n_1,n_2)$ and $x_r(n_1,n_2)$.

Furthermore, the above-derived formula for $x_l$ is limited to values of the
parameters for which the quantities $M$ and $M_+$ are positive. At $M=0$, the
two roots $x_l$ and $x_c$ coincide. In Fig.~\ref{fig1}, this corresponds to the
maximum of the $q$-curve lying on the $x$-axis.  For $M<0$, the maximum of the
curve is below the $x$-axis and the two roots become nonreal.

\subsection{Acceleration}
\label{appb2} Solitons separation $\Delta=x_r-x_l$ is calculated using
Eqs.~(\ref{xreq}) and (\ref{xleq}). This gives rise to Eq.~(\ref{deltaeq}) with
coefficients
\begin{equation}
{\beta_1}=\frac{4 {n_s} {Y_r}^{\frac{{n_s}}{{n_d}+{n_s}}}}{({n_d}+{n_s})
   \left(g_0^2 {n_d}^2 e^{2 \gamma_0}+4 {\eta_d}^2\right)},
\end{equation}
\begin{equation}
{\beta_2}=\frac{4 g_0 ({Y_+}+1)^2 \left(g_0^2 {n_d} {n_s} e^{2 \gamma_0}
   ({n_d}+{n_s} {Y_+})+4 {\eta_d}^2 ({n_d} {Y_+}+{n_s})\right)}{\sqrt{M}
   \left(g_0^2 e^{2 \gamma_0} ({n_d}+{n_s} {Y_+})^2
   +4 ({Y_+}+1)^2 {\eta_d}^2\right)},
\end{equation}
\begin{equation}
{\beta_3}=\frac{8 ({Y_+}+1)^2 {\eta_d} \left(g_0^2 e^{2 \gamma_0}
\left({n_d}^2+{n_s}^2 {Y_+}\right)+4 ({Y_+}+1) {\eta_d}^2\right)}{\sqrt{M}
\left(g_0^2 e^{2 \gamma_0}
   ({n_d}+{n_s} {Y_+})^2+4 ({Y_+}+1)^2 {\eta_d}^2\right)},
\end{equation}
\begin{equation}
{\beta_4}=-\frac{8 {\eta_d} {Y_r}^{\frac{{n_s}}{{n_d}+{n_s}}}}{g_0
({n_d}+{n_s})
   \left(g_0^2 {n_d}^2 e^{2 \gamma_0}+4 {\eta_d}^2\right)}.
\end{equation}

Noting that:\\
\begin{math}
{{\dot x}_d}(t)={\eta_d} e^{\gamma (t)-\gamma_0}-{x_d}(t) {\dot\gamma}(t),
\end{math}\\
\begin{math}
{{\ddot x}_d}(t)={x_d}(t) \left({\dot\gamma}(t)^2-{\ddot\gamma}(t)\right),
\end{math}\\
\begin{math}
{\dot z}_l(t)=-\frac{1}{8} e^{2 \gamma (t)-2 \gamma_0} \left(g_0^2 n_d n_s e^{2
\gamma_0}-4
   {\eta_d}^2\right),
\end{math}\\
\begin{math}
{\ddot z}_l(t)=-\frac{1}{4} e^{2 \gamma (t)-2 \gamma_0} {\dot\gamma}(t)
\left(g_0^2 n_d n_s e^{2 \gamma
   (0)}-4 {\eta_d}^2\right),
\end{math}\\
\begin{math}
{\dot z}_r(t)=-\frac{1}{8} e^{2 \gamma (t)-2 \gamma_0} \left(g_0^2 n_d n_s e^{2
\gamma_0}+4
   {\eta_d}^2\right),
\end{math}\\
\begin{math}
{\ddot z}_r(t)=-\frac{1}{4} e^{2 \gamma (t)-2 \gamma_0} {\dot\gamma}(t)
\left(g_0^2 n_d n_s e^{2 \gamma
   (0)}+4 {\eta_d}^2\right)
\end{math},\\\\
the acceleration, ${\ddot\Delta}(t)$, can be calculated to take the form of
Eq.~(\ref{force2}) with coefficients
\begin{eqnarray}
\beta_5&=& \frac{{ Y_+}^{\frac{n_d+n_s}{2 n_d-2 n_s}}} {64
\sqrt{{Y_r}}}e^{\left(-\frac{1}{4} g_0 \Delta
   n_d e^{\gamma (t)}-5 \gamma_0\right)}
   \left(4 g_0 e^{2 \gamma_0} {\eta_d}^2 \left(g_0 {\beta_3}
   \left(n_d^2+4 n_d n_s+n_s^2\right)+4 {\beta_2}
   {\eta_d} (n_d+n_s)\right)\right.\nonumber\\&-&\nonumber \left.
   g_0^3 n_d n_s e^{4 \gamma_0} \left(g_0 n_d
   n_s {\beta_3}+4 {\beta_2} {\eta_d} (n_d+n_s)\right)-16 {\beta_3}
   {\eta_d}^4\right),
\end{eqnarray}
\begin{eqnarray}
\beta_6&=& -\frac{{Y_r}^{\frac{n_d}{n_d+n_s}-\frac{1}{2}}}{64 \sqrt{{ Y_+}}}
e^{ \left(\frac{1}{4} g_0 \Delta
   n_d e^{\gamma (t)}-3 \gamma_0\right)}
   \left(-4 g_0 e^{2 \gamma_0} {\eta_d}^2 \left(g_0 {\beta_4}
   \left(n_d^2-4 n_d n_s+n_s^2\right)+4 {\beta_1}
   {\eta_d} (n_d-n_s)\right)\right.\nonumber\\&+&\nonumber \left.g_0^3 n_d n_s e^{4 \gamma_0} \left(g_0 n_d
   n_s {\beta_4}+4 {\beta_1} {\eta_d} (n_s-n_d)\right)+16 {\beta_4}
   {\eta_d}^4\right),
\end{eqnarray}
\begin{eqnarray}
\beta_{7}&=&\frac{{ Y_+}^{\frac{n_d+n_s}{2 n_d-2 n_s}}}{64 \sqrt{{Y_r}}}
e^{\left(-\frac{1}{4} g_0 \Delta
   n_d e^{\gamma (t)}-4 \gamma_0\right)} \left(4 g_0^2 e^{2 \gamma_0}
   {\eta_d} \left(g_0 n_d n_s {\beta_3} (n_d+n_s)+{\beta_2} {\eta_d}
   \left(n_d^2+4 n_d n_s+n_s^2\right)\right)\right.\nonumber\\&+&\nonumber \left.g_0^4 n_d^2 n_s^2 {\beta_2}
   \left(-e^{4 \gamma_0}\right)-16 {\eta_d}^3 \left(g_0 {\beta_3} (n_d+n_s)
   +{\beta_2} {\eta_d}\right)\right),
\end{eqnarray}
\begin{eqnarray}
\beta_8&=&-\frac{{Y_r}^{\frac{n_d}{n_d+n_s}-\frac{1}{2}}}{64 \sqrt{{ Y_+}}}
e^{\left(\frac{1}{4} g_0 \Delta
   n_d e^{\gamma (t)}-2 \gamma_0\right)} \left(-4 g_0^2 e^{2 \gamma_0}
   {\eta_d} \left(g_0 n_d n_s {\beta_4} (n_s-n_d)+{\beta_1} {\eta_d}
   \left(n_d^2-4 n_d n_s+n_s^2\right)\right)\right.\nonumber\\&+&\nonumber \left.
   g_0^4 n_d^2 n_s^2 {\beta_1}
   e^{4 \gamma_0}+16 {\eta_d}^3 \left(g_0 {\beta_4} (n_d-n_s)
   +{\beta_1} {\eta_d}\right)\right).
\end{eqnarray}

\newpage

\end{document}